\documentclass[fleqn,usenatbib]{mnras}

\usepackage{newtxtext,newtxmath}

\usepackage[T1]{fontenc}
\usepackage{ae,aecompl}

\usepackage{graphicx}	
\usepackage{xcolor}

\title[Cosmic Filaments Delay Quenching Inside Clusters]{Cosmic Filaments Delay Quenching Inside Clusters}

\author[S. Kotecha et al.]{
Sachin Kotecha,$^{1}$
Charlotte Welker,$^{1,2}$\thanks{E-mail: cwelker3@jh.edu}
Zihan Zhou,$^{1,3}$
James Wadsley,$^{1}$
\newauthor{
Katarina Kraljic,$^{5}$
Jenny Sorce,$^{6}$
Elena Rasia,$^{7}$
Ian Roberts,$^{1}$
}
\newauthor{
Meghan Gray,$^{4}$
Gustavo Yepes,$^{8}$
Weiguang Cui$^{5}$
}
\\
$^{1}$ Department of Physics and Astronomy, McMaster University, Hamilton, Ontario, Canada\\
$^{2}$ Department of Physics and Astronomy, The Johns Hopkins University, Baltimore, Maryland, USA\\
$^{3}$ Department of Astronomy, School of Physical Sciences, University of Science and Technology of China, Hefei, Anhui 230026, China\\
$^{4}$ Department of Astronomy, University of Nottingham, UK\\
$^{5}$ Department of Physics and Astronomy, University of Edinburgh, Edinburgh, UK\\
$^{6}$ Centre de Recherche Astrophysique de Lyon, Universite de Lyon, Lyon, France\\
$^{7}$ Osservatorio Astronomico di Trieste, Trieste, Italy\\
$^{8}$ Departamento de Fisica Teorica, Modulo 15, Facultad de Ciencias, Universidad Autonoma de Madrid, 28049 Madrid, Spain\\
}

\date{Accepted XXX. Received YYY; in original form ZZZ}

\pubyear{2020}

\begin{document}
\label{firstpage}
\pagerange{\pageref{firstpage}--\pageref{lastpage}}
\maketitle

\begin{abstract}
We investigate how large-scale cosmic filaments impact the quenching of galaxies within one virial radius of 324 simulated clusters from {\bf The Three Hundred} project. We track cosmic filaments with the versatile, observation-friendly program DisPerSE and identify halos hosting galaxies with VELOCIRaptor. We find that cluster galaxies close to filaments tend to be more star-forming, bluer, and contain more cold gas than their counterparts further away from filaments. This effect is recovered at all stellar masses. This is in stark contrast with galaxies residing outside of clusters, where galaxies close to filaments show clear signs of density related pre-processing. We first show that the density contrast of filaments is reduced inside the intra-cluster medium. Moreover, examination of flows around and into cluster galaxies shows that the gas flows in intra-cluster filaments are colder and tend to stream along with galaxies in their midst, partially shielding them from strangulation by the hot, dense intra-cluster medium. This also preserves accretion onto satellites and limit ram pressure. 

\end{abstract}




\section{Introduction}
\label{sec:introduction}

On the largest scales of the Universe, matter is distributed in a complex structured network visible in the galaxy field from spectroscopic surveys \citep{deLapparent86,Geller89, colless,Stoughton_2002,Doro04}. This large-scale structure, often referred to as `the cosmic web' \citep{Bond_1996}, naturally arises from the gravitational collapse of initial perturbations in the density field of the early Universe \citep{peebles,Zeldo70,Shandarin89,Hidding14}. We observe large under-dense regions, or voids. These voids are segmented by sheet-like regions of higher density known as walls, which attract matter from voids. In turn, these walls are further collapsing into a network of dense filaments that delineate them. These filaments channel matter towards the highest density regions of the Universe, the cosmic nodes, where they meet. It is at the nodes of this cosmic web that we find galaxy clusters, with filaments diving deep into their centres. Galaxy clusters are the densest and most massive gravitationally-bound structures in the Universe, hosting thousands of galaxies. They are therefore a hotspot for both galaxy interactions and complex non-linear dynamics. In particular, the intra-cluster medium is a place of increased galaxy quenching: the rapid shutting down of star formation in galaxies evolving in dense environments \citep[][and references therein]{Brown17}.

It is well-established that the star formation activity of galaxies in today's Universe distributes in a clear bimodal fashion, easily detected in observations through the corresponding colour bimodality of the light they emit  \citep{Strateva_2001, balogh, Baldry_2004}. Bluer galaxies are home to younger, hotter stars and are hence more actively star forming, while their reddest peers no longer form stars, and are effectively `dead'. Between these two populations, few galaxies are found to display intermediate star formation patterns, suggesting that galaxies undergo a rapid shutting down of their star formation at some point in their history \citep{Salim14}. A perennial question of galaxy evolution is therefore exactly how galaxies turn off their star formation, or quench. 

It is well understood that stars form when cold, dense gas collapses, and as such, star formation shuts off when cold gas supply in or around a galaxy becomes scarce \citep{gabor}. This quenching occurs preferentially in dense environments such as clusters and possibly involves a variety of processes. For instance, strangulation occurs when there is no more gas inflow to a galaxy (e.g. \citep{larson, balogh_2000, Peng_2015}. Stars continue to form using up the available gas until there is little to none left, at which point a galaxy is quenched.  Harassment and mergers occur due to interactions between galaxies, whereby gas can be heated and/or stripped, causing star formation to shut down in such galaxies \citep{Moore_1996, smith}.

The radial trends resulting from the progressive quenching of galaxies falling into clusters have been extensively described in both simulations and observations: it is now established that the fraction of star-forming galaxies decreases towards the centre of clusters as quenching proceeds  \citep[e.g.][]{balogh_2000, haines, Raichoor_2012, Pintos_Castro_2019}.

In recent years, increasingly robust computational models and capabilities have allowed more accurate and effective simulations of the complex environments of galaxy clusters and improved understanding of how galaxies evolve in these regions \citep{Barnes_2017, cui, tremmel}.  Among other mechanisms, ram pressure stripping seems to play a major role deep into clusters \citep[e.g.][]{Wetzel_2013, roberts, maier2019slow, Muzzin14, Peng_2015,Brown17}. We define ram pressure stripping as the interaction of galaxies with the hot gas they are moving through \citep{gunngott}. 


Interestingly, outside of clusters, some recent studies suggest cosmic filaments might also play a role in the early quenching of field galaxies at low redshift. Indeed, in the field, large cosmic filaments are denser than their surroundings, and as such may be associated with pre-processing \citep{kraljic, Laigle_2017, Sarron_2019}. While galaxies near such filaments are expected to display higher stellar masses on average \citep{welker}, they also undergo  more interactions, and thus may quench faster and become redder than their counterparts further away from the filaments \citep{kraljic}.

On the other hand, it has long been theorized that, since gas shocks and cools into cosmic filaments, such filaments should be able to funnel collimated streams of cold gas crucial to star formation deep into halos  \citep{Katz_1993, brooks2009, Keres_2009, kleiner}, at least at $z>1$. To what extent such cold flows might stream deep into clusters, past the virial shock is dependent on factors such as the density contrast of the filament, the degree of turbulence \citep{Cornuault18} or the intensity of feedback processes \citep{Dubois13,Powell_2011}. Nonetheless, even after collimation is lost to hydrodynamic instabilities, and monophasic flows have blended in  the intra-cluster medium (ICM), relics of the cooler core of large inter-cluster filaments  might still locally pervade the ambient medium \citep{Mandelker19,Mandelker20}. It is therefore conceivable that filaments plunging into clusters  might linger in diffuse forms even at $z<1$, possibly modifying the local dynamics of the ICM and impacting the evolution of local satellite galaxies. 

However, the impact of large inter-cluster cosmic filaments on quenching deep into clusters (`intra-cluster' filaments) remains seldom analysed. Therefore, their role on local star formation remains to be explored. 

This is in part because most cosmic web extractors either do not operate on the gas field \citep{Tempel16}, or identify cosmic structures as either a filament or a cluster, so they do not allow for the analysis of hybrid structures such as intra-cluster filaments \citep{Cautun15}. To bridge this gap, we use DisPerSE \citep{sousbie}, a versatile topological extractor that identifies the spine of the cosmic web and therefore allows us to track filaments down to the core of clusters. 

The lack of analysis is also because large gaseous filaments remain elusive in observations. However, recently, cosmic gas filaments were directly observed for the first time stretching between galaxies in the SSA22 protocluster at $z=3.1$ with the Multi Unit Spectroscopic Explorer on the Very Large Telescope \citep{umehata}. Upcoming facilities such as the SKA might shed even more light on intra-cluster filaments in the near future \citep{Locatelli18}.

The focus of this study is therefore to analyze the impact of intra-cluster cosmic filaments on quenching, that is, in regions of the universe where the background is a dense hot intra-cluster medium and where gas stripping becomes one of the dominant quenching mechanisms. In this work we use a robust set of hydrodynamic cluster simulations to analyse the effects of cosmic filaments on galaxy evolution inside the cluster environment, and in particular determine whether somewhat cooler gas streams are present near filaments at $z=0$ and are able to delay quenching locally.

The remainder of this paper is structured as follows. In Section~\ref{sec:nummeth} we discuss the data products and numerical quantities used for this work. In Section~\ref{sec:quenching} we analyze the quenching of galaxies around cosmic filaments in different environments through a variety of tracers. In Section~\ref{sec:dynamics} we look at the dynamics of gas and halos surrounding filaments themselves to understand the physical origin of these quenching trends. Finally, in Section~\ref{sec:summary} we summarize our findings and discuss next steps.

\section{Numerical Methods}
\label{sec:nummeth}
In this section, we describe the galaxy cluster simulation we used in ~\ref{sec:the300}, the halo finder in ~\ref{sec:velociraptor}, and the filament extractor in ~\ref{sec:disperse}. We then discuss the tracers we use to investigate the evolution of star formation (\ref{sec:quetrac}) and the dynamics of gas and galaxies (\ref{sec:dynmeas}).

\subsection{Simulation Sample: The Three Hundred}
\label{sec:the300}

A large sample of galaxy cluster simulations with sufficient resolution is required in order to have robust statistics on intra-cluster filaments and satellite galaxies. For this study, we use The Three Hundred project\footnote{https://www.the300-project.org} \citep{cui}, a set of 324 simulated galaxy clusters run with a smoothed-particle hydrodynamic (SPH) scheme.

The clusters in this set are hydrodynamic zoom re-simulations of the most massive halos  in the dark matter-only  MultiDark MDPL2 simulation \citep{klypin} at $z=0$, run with GADGET-X, a version of GADGET3 with the improved SPH scheme of \citet{beck} \citep[itself an updated version of the GADGET2 code,][]{springel}. 

The Three Hundred suite was run with cosmological parameters $\Omega_M=0.307,\Omega_B=0.048,\Omega_{\Lambda}=0.693,h=0.678$, a dark matter particle mass of $M^{0}_{DM}=1.81\times 10^9 M_{\odot}$, and gas particle mass of $M^{0}_{gas}=3.39\times 10^8 M_{\odot}$, corresponding to a typical spatial resolution of 8 kpc.$h^{-1}$. It includes advanced prescriptions for processes such as metal-dependent UV cooling, black hole growth and AGN feedback, star formation and stellar evolution, and stellar feedback with varying star particle masses \citep{cui}.


Clusters range from $M_{\rm 200} = 4\times10^{14}M_{\odot}$ to $10^{15}M_{\odot}$, and vary from  relaxed to unrelaxed. We conduct the study at $z=0$. In order to avoid boundary effects at the edge of the re-simulated regions, we limit our analysis to a 14 $h^{-1}$.Mpc spherical region from the centre of each cluster. In practice, since most of our analysis focuses on halos within 2 virial radii ($R_{\rm vir}$) of the clusters, our results are not affected by this cut.  

With varying star particle masses on the order of $M_{\rm star}=5\times10^7 M_{\odot}$, galaxies with more than 50 particles are resolved down to stellar masses of $10^{9.5}M_{\odot}$.

\subsection{Halo identification with VELOCIraptor}
\label{sec:velociraptor}

Halos hosting galaxies were identified with the VELOCIraptor halo finder \citep{elahi}. We use the mixed-component version of the finder, which operates on dark matter (DM), gas, and star particles. The finder conducts a multi-step algorithm to identify structures. It begins by identifying halos with a 3D friends-of-friends (FOF) first pass search, followed by a 6D FOF search in phase space. The process allows us to identify kinematically coherent substructures and mergers. Stellar mass, virial radii, and virial masses are computed on the fly following \cite{BryanNorman98}. Note that for galaxy clusters $R_{\rm vir}\approx R_{100}\approx 1.4 \times R_{\rm 200}$. For this reason, we adopt the definition 
$R_{\rm vir} = R_{100}$ for clusters in the present study. It is also physically motivated by \cite{Arthur19} who found the main gas accretion shock in the clusters used here to be around $\approx 1.5 \, R_{200}$.
In our sample, the clusters virial radii $R_{\rm 100}$ therefore range from 2 to 6 $h^{-1}$.Mpc.

Note that this mixed-component finder is particularly efficient at identifying halos and sub-halos hosting galaxies rather than galaxies on their own. As a consequence, a single, intermediate size halo containing an advanced galaxy merger with no remaining identifiable DM sub-halo will appear as a single structure rather than two galaxies in the process of merging. This is however suitable for the limited resolution of the simulation that does not allow for a robust separation of distinct galaxies late in the process of merging across the entire stellar mass range analysed. Our study is therefore conducted at the halo/sub-halo level. 

To ensure that a halo is sufficiently resolved and does contain a galaxy candidate, we exclude any halo with less than 100 DM particles or with a stellar mass $M_{\rm star}<10^{9.5}$M$_{\odot}$. We therefore focus on the gas and stellar content {\it within each satellite halo's virial radius $R_{\rm vir}^{\rm sat}$}. In practice, the cold gas and stellar content are however concentrated at the centre of halos, on galactic scales, and can therefore be assimilated to a galaxy. It follows that, in our study, the terms {\it halo} and {\it galaxy} refer to the same structure in the VELOCIraptor catalog, hence are mostly interchangeable. The first one puts the emphasis on the extended DM+gas component while the second puts the emphasis on the concentrated stellar+cold gas component.

We exclude the brightest cluster galaxies (BCGs), identified as the halo with the highest stellar mass within 0.5 $R_{\rm vir}$ each cluster. This work focuses on the fate of satellite halos/galaxies - defined as all the halos and sub-halos other than the main cluster - falling into a cluster environment either along filaments or on deviated orbits. Infall mechanisms do not directly govern the evolution of BCGs, which seem however sensitive to the connectivity of the cosmic web \citep{kraljiccon}.

\subsection{Filament extraction with DisPerSE}
\label{sec:disperse}
The DisPerSE persistent structure finder \citep{sousbie} is used to identify filaments diving deep into our simulated galaxy clusters. DisPerSE has been successfully used on both simulations  \citep{Laigle_2014, codis, Kraljic_2018,Rost21,Kuchner21} and observational studies  \citep{Laigle_2017,Malavasi17,bird,welker}.

DisPerSE is based on topology rather than geometry, as one of its main advantages. It characterizes topological features (minima, maxima, saddle points, and ridge lines connecting them) as the various components of the cosmic web, including walls, voids, and filaments. Ridge lines of the input density field are returned as a contiguous network of filaments. Further trimming is based on persistence, the density ratio in a pair of linked critical points (such as maximum-saddle point). The persistence threshold translates to a minimum signal-to-noise ratio in this context. These techniques make DisPerSE excellent in the context of large-scale structures, with noisy data, and inherent scale-invariant nature.

For this analysis, DisPerSE was applied to the gas density field of the simulations, first projected on a 30 $h^{-1}$.Mpc wide 3D cubic grid centred on each cluster, with pixels of 150 $h^{-1}$.kpc on each side. It was smoothed over 8 pixels with a Gaussian kernel to focus on the larger cosmic filaments, by effectively getting rid of thinner filaments between large satellites, which otherwise stand out as contrasted peaks in the density field.

\begin{figure*}
	\includegraphics[width=\columnwidth]{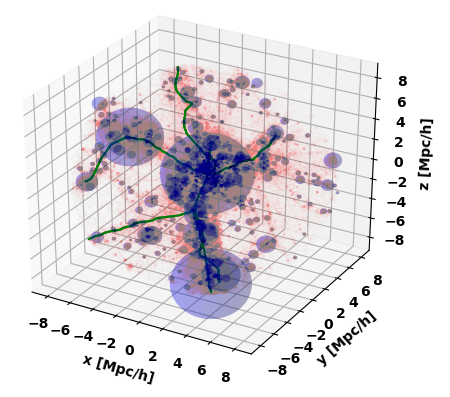}
	\includegraphics[width=\columnwidth]{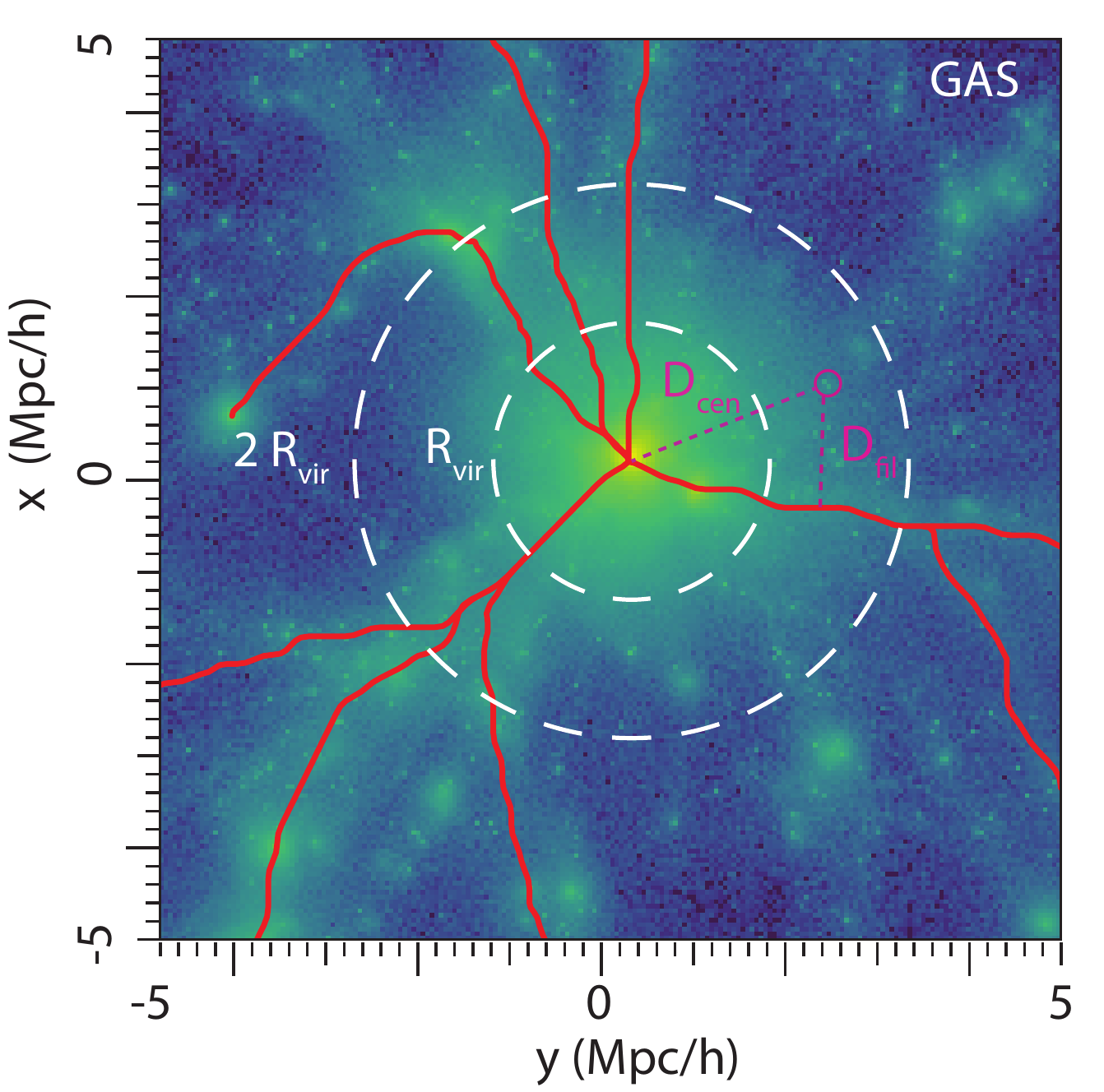}
    \caption{\textbf{Left panel}: 3D map of cluster 50, with filaments in green, dark matter in pink, and halos virial radii as blue circles. \textbf{Right panel}: 2D z-projected map of the gas density in Cluster 50. Filaments are overplotted in red and the cluster's virial radius is indicated as white dashed circles. $D_{\rm fil}$ and $D_{\rm cen}$ for a given satellite appear in pink.}
    \label{fig:casca_diagram_zoomout}
\end{figure*}

As an example, a visual representation of Cluster 50 is shown in Fig.~\ref{fig:casca_diagram_zoomout}. The left panel depicts a 3D representation with filaments in green, galaxy halos defined by their virial radius in blue, and dark matter particles in pink. The cluster's virial radius appears as the large central blue sphere. Two infalling groups are visible, and numerous galaxies clearly align along the spine of the cosmic web. The right panel of Fig.~\ref{fig:casca_diagram_zoomout} displays a zoom-in of the same cluster projected along the z axis. The color palette varies linearly with the log of the gas density. Projected filaments are in red and 1 and 2 $R_{\rm vir}$ are indicated as white dashed circles.  One can clearly see the gas filaments diving deep into the cluster, with gas-rich haloes along them. We demonstrate in Appendix~\ref{sec:mstevolapp} how the most massive halos reside along the spine of filaments.

For each satellite halo within 10 $h^{-1}$.Mpc of its host cluster, we further define $D_{\rm fil}$, the distance of a halo to its nearest cosmic filament, and $D_{\rm cent}$, its distance to the cluster centre (in pink on the right panel of Fig.~\ref{fig:casca_diagram_zoomout}).  
The reader will notice that $D_{\rm fil}$ and $D_{\rm cent}$ are not by construction independent coordinates as they are not orthogonal. We therefore systematically compute the ratio ${D_{\rm fil}}/{D_{\rm cent}}$, which is in essence the separation angle of the satellite halo with respect to its nearest filament, as seen from the centre of the host cluster. Throughout this work we call this ratio the `angular separation'. This quantity allows us to isolate the evolution of halo properties driven specifically by the proximity to a filament and not simply by the progressive infalling into a dense cluster. 

Note that one has two choices for $D_{\rm cent}$: it can be computed as the distance to the cluster's centre of mass (DM+gas+stars, dominated by DM+stars), or as the distance to the node of the gaseous cosmic web corresponding to the halo, which is effectively the peak of the cluster's gas distribution. The difference between centers is typically $<10\%\,R_{\rm vir}$. Major results are qualitatively similar for both methods, we therefore focus on the first in the present study, as it is the closest to observations which typically use the BCG as the cluster's centre. A direct consequence of this choice is that, in the inner core of the cluster ($0.3 R_{\rm vir}$), it is possible to have $D_{\rm fil}>D_{\rm cent}$ for up to $10\%$ of galaxies. Note also that some particularities arise for the most unrelaxed clusters. They are the subject of a follow-up paper (Welker et al. 2021, in prep).

\subsection{Quenching Tracers}
\label{sec:quetrac}

In the following section, we outline the various tracers and observables we designed to investigate quenching in this study.

\subsubsection{Star Formation Rate}
\label{sec:sfr}
The most direct property to examine when it comes to quenching is the star formation rate (SFR), the number of stars a galaxy is birthing per year, or the specific star formation rate (sSFR), such that $sSFR=SFR/M_{\rm star}$.

We use two avenues to compute SFR values in The Three Hundred. In the {\it instantaneous} method, we identify gas particles within $R_{\rm vir}^{\rm sat}$ of each satellite halo and sum up SFR values in $M_{\odot}/yr$ assigned to gas particles by the GADGET-X subgrid star-formation recipe. This typically identifies new stars within a 50 Myr age range.
 
In the {\it secular} method, we sum up the masses of the star particles that have formed within the past 1 Gyr (based on their age) within one virial radius ($R_{\rm vir}^{\rm sat}$) of the satellite halo. This method is expected to be less sensitive to short-lived interactions and to the limited resolution of star formation in the simulation (mainly because of the long time interval chosen).
 
However, owing to the limited resolution of the simulation itself, the subgrid star formation recipe of the simulation is also low resolution and suffers discrete numerical effects that affect both methods. Low mass galaxies can easily appear to not be star forming due to the stochasticity of the star formation model.  On the contrary, massive galaxies do not stop star formation early or fast enough as a result of runaway star formation by the subgrid model that fails to self-regulate in densest environments, and weak AGN and stellar feedback. 

As a result, both simulated SFRs trend strongly and log-linearly with stellar mass across the full mass range, including for galaxies with $M_{\rm star}>10^{11.5} M_{\odot}$, past the point where the trend is expected to reverse. This suggests that quenching evaluated from SFRs will be significantly underestimated at high mass due to this numerical artifact. 

Mass rankings are however preserved and the analysis of sSFR at fixed stellar mass can still be used to obtain reliable qualitative results. To this end, we restrict the analysis to the fraction of star-forming galaxies in a given bin of $N$ galaxies, defined as:

\begin{equation}
    f_{\rm SFG}= \frac{n(\rm sSFR>sSFR_{0})}{N}
	\label{eq:SFG}
\end{equation}
\noindent with $N$ the total number of galaxies in the bin considered and $n(\rm sSFR>sSFR_{0})$ the number of galaxies with $\rm sSFR>sSFR_{0}=10^{-11}$ yr$^{-1}$, a typical value used for such a cut \citep{Wetzel_2013, roberts}.
We use bi-dimensional bins of stellar mass and property of interest (such as distance to filament).
 
There is nevertheless a need for less direct but more robust tracers of quenching, which we detail in the following sections.
 
\subsubsection{Galaxy Colours}
\label{sec:galcols}

Galaxy colours are directly tied to star formation but are more related to the specific stellar make-up of the galaxy than to its mass, and are integrated over its whole history, thus less sensitive to aforementioned resolution limits. They also allow for direct observational comparisons. As galaxies quench, their stellar content ages without being refreshed by new stars, hence galaxies become redder. 

We therefore compute galaxy SDSS colours for all our populated halos using stellar population models. We follow the method detailed in \citet{Dubois14}, whereby we apply a Single-Burst Stellar Population model (SSP) from \citet{Bruzual_2003} to the star particles in the simulation using the metallicity, age, and mass (obtained using the Salpeter initial mass function) of the particles to obtain the flux per frequency. The contributions from all the stars within  $R_{\rm vir}^{\rm sat}$ are passed through the various SDSS filters. Note that in this simple model, we do not take into account any extinction from dust. From this, we obtain stellar luminosities which we mass average to assign luminosities to the galaxies. We then use these luminosities to obtain the galaxy colours. In this work, results are presented for $g-r$. Based on the simulated colour bimodality, a typical red galaxy in our simulation is one with $g-r\geq0.65$. This is consistent with the threshold from low-z SDSS galaxy populations \footnote{https://www.sdss.org/}.

\subsubsection{Gas Fraction}
\label{sec:gasfrac}
Gas content, particularly cold gas content, is the fuel for star formation and therefore reliably traces the ability of halos to form stars. 

In The Three Hundred, all halos with $M_{\rm tot}>10^{10.5} M_{\odot}$ have a gas-to-total initial mass fraction resolved with at least 0.1\% precision. Note however that, at our resolution, pressure is resolved only down to a lower threshold $P_{\rm floor}$, resulting in a typical lower limit of $10^4$ K for temperatures in halos. This is comparable to other flagship kiloparsec scale cosmological simulations such as Horizon-AGN, Eagle or IllustrisTNG \citep{Dubois14, Schaye_2014, Crain_2015, Nelson_2018}. In particular, we do not resolve the formation of molecular clouds.

We therefore identify as cold halo gas any gas particle with a temperature at or below $10^5$ K within 1 $R_{\rm vir}^{\rm sat}$ of a satellite halo. Based on the cooling model used in The Three Hundred clusters and identification of the radiative cooling branch on the temperature-density diagrams, gas in this temperature range trapped within simulated halos corresponds to gas that would typically cool down into cold HI and the ISM in real galaxies \citep[see][]{Rost21}. 
 
We define the cold gas fraction of the galaxy-hosting halos in our sample as:

\begin{equation}
    f_{\rm gas}^{\rm cold}=\frac{M_{\text{gas}}(\text{T}\leq10^5\text{ K})}{M_{\text{gas}}(\text{T}\leq10^5\text{ K})+M_{\rm star}}
	\label{eq:CGF}
\end{equation}

\noindent with $M_{\rm gas}(\rm T)$ the total mass of the gas with temperature below T within 1 $R_{\rm vir}^{\rm sat}$ of the halo, and $M_{\rm star}$ its total stellar mass within 1 $R_{\rm vir}^{\rm sat}$.

\subsection{Dynamical Measures}
\label{sec:dynmeas}
To analyse the origin of satellite quenching in clusters, we also study the dynamics of these galaxies through the gaseous medium surrounding them. This section outlines the quantities we define to investigate these dynamics.

\subsubsection{Filament Kinematics}
\label{sec:flows}
In order to investigate how coherent flows of gas and galaxies are along cosmic filaments, we measure the angle between the velocity vector of halos and gas particles with respect to their nearest filament. More specifically, for every satellite halo or gas particle with cluster-centric velocity, $\vec{v}$, and its nearest filament segment,  $\vec{f}_{near}$, oriented inwards, we compute:

\begin{equation}
\cos{\theta}=\frac{\vec{f}_{\rm near}\cdot\vec{v}}{|\vec{f}_{\rm near}||\vec{v}|}.
\label{eq:cos}
\end{equation}

By definition, filament segments are increasingly radial towards the cluster centre, which is a node of the cosmic web. Motions of galaxies and gas accreted into clusters are also expected to be radial on average, with various degrees of dispersion depending on proximity to the cluster centre.

Thus, even in the absence of any specific filamentary kinematic enhancement, the correlation between $\vec{v}$ and $\vec{f}_{near}$ is expected to increase with decreasing $D_{\rm fil}$ and $D_{\rm fil}/D_{\rm cen}$. 

In order to separate a potential filament boost from this purely geometrical effect, we compute the null hypothesis (pure geometrical correlation only) by calculating $\cos{\theta}_{\rm H_{\rm 0}}$ after rotating the filament network of each cluster by $45^{\circ}$ in a random direction. Any filament-specific correlation with $D_{\rm fil}$ is destroyed in the flip, whereas purely geometric effects should be left unchanged on average by this procedure. The choice of $45^{\circ}$ is justified by the fact that the average connectivity of our clusters (the number of connecting inter-cluster filaments globally) is around 6, hence the average angle between 2 connecting filaments is close to $90^{\circ}$.

\subsubsection{Accreting Fraction}
\label{sec:accfrac}
The inability of a halo to accrete and/or retain cold gas, often referred to as strangulation, is key to its environmental quenching. We compute the inflow/outflow into/from every satellite halo individually. We first identify cold ($T<10^5K$) gas particles within a spherical shell with thickness $\Delta r=0.1$ $h^{-1}$.Mpc, at 0.5 $R_{\rm vir}^{\rm sat}$ of the halo. This ensures that the gas is indeed associated with the halo. The net outflow is then computed as:

\begin{equation}
\dot{M}_{\rm gas}^{\rm cold}=\frac{\sum_{\rm shell} m_i\vec{v}_i\cdot\hat{r}}{\Delta r}
\label{eq:outflow}
\end{equation}
\noindent 
with $m_{i}$ the mass and $\vec{v}_i\cdot\hat{r}$ the halo-centric outward radial velocity of gas particle $i$.

Note that a negative outflow is actually a net inflow, in which case the corresponding halo is accreting cold gas. 

This allows us to define the accreting fraction of a given sample of $N$ satellite halos as:

\begin{equation}
f_{\rm acc}=\frac{n(\dot{M}_{\rm gas}^{\rm cold}<0)}{N}
\label{eq:accfrac}
\end{equation}

\subsubsection{Ram Pressure Stripping}
\label{sec:rps}
Ram pressure stripping is expected to be the dominant quenching mechanism in the core of clusters \citep{Wetzel_2013, roberts, maier2019slow}. We estimate the ram pressure term on each satellite halo as: 
\begin{equation}
     P_{\rm ram}=\rho_{\rm shell} (\delta v)^2
    \label{eq:rps}
\end{equation}

\noindent with $\delta v$ the halo to intra-cluster medium relative velocity and $\rho_{\rm shell}$ the density of the hot intra-cluster gas that the halo is passing through. $\rho_{\rm shell}$ is estimated as an average of individual densities, weighted by each particle's volume ("volume-weighted"), for all `environment' particles within a 0.35 $h^{-1}$.Mpc (0.5 Mpc) thick spherical shell past the halo's virial radius $R_{\rm vir}^{\rm sat}$. Only gas particles with  $\rm T>10^5$K are considered since colder particles are expected to be escaping or bound to the halo. 

$\delta v$ is defined as $|\vec{v}_{\rm halo}-\vec{v}_{\rm enviro}|$, with $\vec{v}_{\rm enviro}$ the mass-weighted average velocity of hot gas particles in the `environment' shell and $\vec{v}_{\rm halo}$ the mass-weighted average velocity of  all particles (dark matter, star, gas) within 1 $R_{\rm vir}^{\rm sat}$ of the halo.

We expect the stripping due to ram pressure to be somewhat excessive in this simulation, as in most numerical simulations with similar resolutions. Indeed, the inner cores of satellite halos are not resolved with enough particles to properly account for their steep density gradient. As a result, such halos tend to be more easily stripped than their observed counterparts. Note that relative and qualitative differences, rather than exact amounts, are the focus of this work.

\subsubsection{Gas Unbinding Parameter}
\label{sec:bgdef}

To better estimate the individual degree of gas stripping a galaxy-hosting halo undergoes, we construct for each satellite halo an instantaneous gas unbinding parameter, $B_g$, which quantifies the degree of displacement between the gas particles and the stellar and dark matter particles in their phase-space distributions, within 1 $R_{\rm vir}^{\rm sat}$ of the halo, excluding material in the halos's satellites if some are present .  For a given halo, $B_g$ is defined as:

\begin{equation}
        B_g=\frac{\sqrt{(v_{x,\rm g}-v_{x,\rm sdm})^2+(v_{y,\rm g}-v_{y,\rm sdm})^2+(v_{z,\rm g}-v_{z, \rm sdm})^2}}{\sqrt{\sigma_{v_x,\rm s}^2+\sigma_{v_y,\rm s}^2+\sigma_{v_z,\rm s}^2}}
	    \label{eq:Bg}
\end{equation}

\noindent where $\vec{v}_{\rm g}$, $\vec{v}_{\rm s}$ and $\vec{v}_{\rm sdm}$  are the mean velocities of the gaseous, stellar, and stellar$+$dark matter components of the halo respectively, computed within 1 $R_{\rm vir}^{\rm sat}$ as averages over corresponding particles. Star and dark matter particles averages are mass-weighted while gas particles averages are density-weighted to avoid biases due to high-velocity, fly-through gas particles from the hot but less dense ICM. In this study, ICM refers to all gas particles with $T>10^{5} K$ within the cluster's virial, radius not bound to a satellite. $\sigma_{v,s}$ is the stellar velocity dispersion within 1 $R_{\rm vir}^{\rm sat}$.

Note that we do not impose a cut on gas temperature here. That way, most galaxies devoid of cold gas can still be assigned a $B_g$ values, which is indicative of the velocity of their DM and stellar components with respect to the ICM gas, hence of the stripping capacity of their direct environment.

\begin{figure}
	\includegraphics[width=\columnwidth]{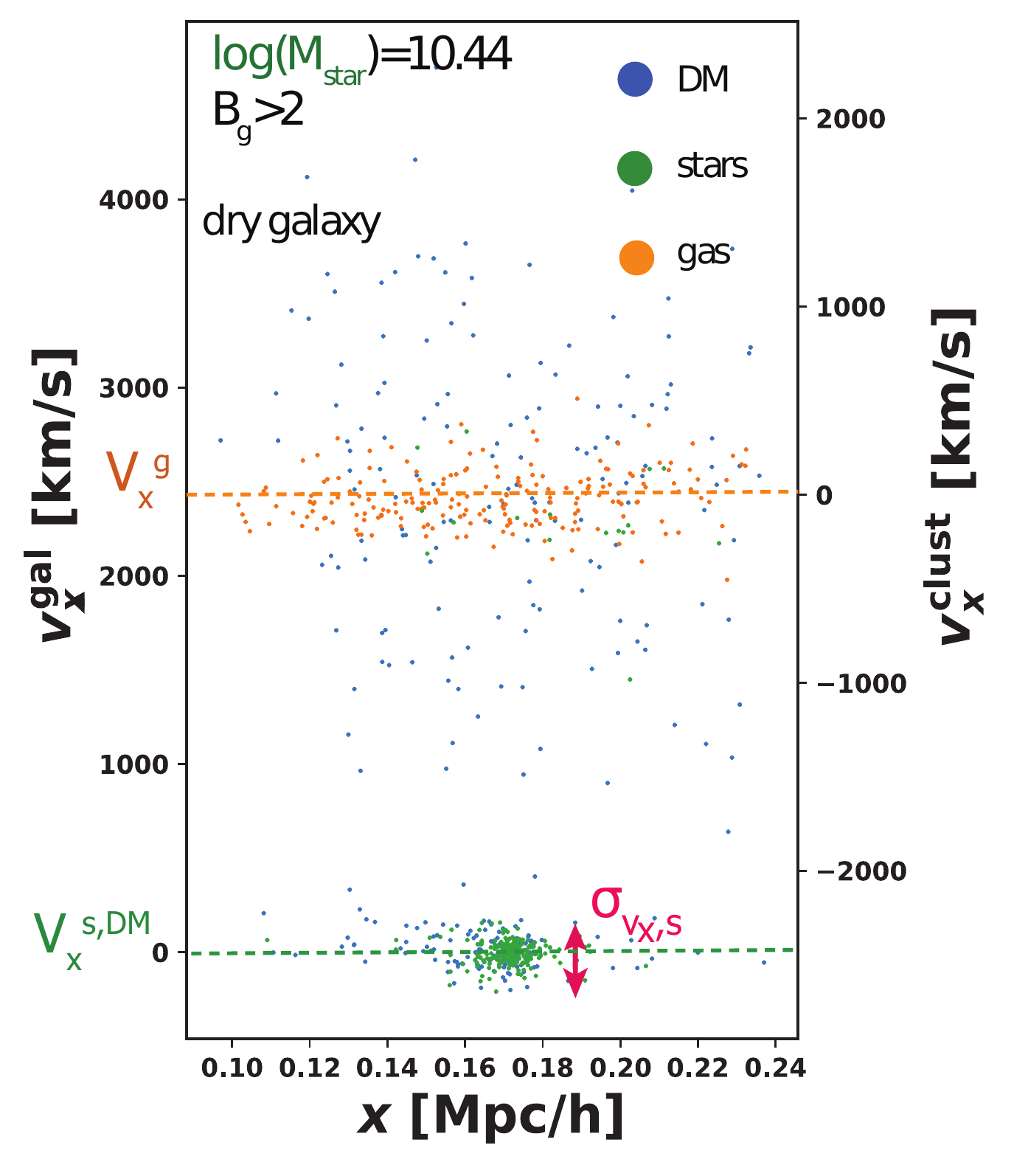}
    \caption{Illustration of the construction of $B_g$ on the 1-D phase-space diagram of a dry satellite galaxy. Velocities are in the frame of the cluster on the right and in that of the galaxy on the left. $B_g$ quantifies the velocity offset of gas (orange) from the stellar (green) and dark matter (blue) components of the halo, normalized by the stellar velocity dispersion (in red).}
    \label{fig:gasbindexplan}
\end{figure}

Fig.~\ref{fig:gasbindexplan} displays a 1-D phase space diagram which shows how $B_g$ is constructed on the example of a dry halo (no cold gas) hosting a $M_{\rm star} \approx 10^{10.4} M_{\odot}$ galaxy, illustrated on one Euclidean component of $B_g$ only. The panel shows the $x$ component of the position vector to the centre of the cluster on the horizontal axis and $x$ velocities on the vertical axis (in the frame of the galaxy on the left and the frame of the cluster on the right). Gas particles are plotted in orange, stellar particles in green, and dark matter particles in blue. The gas content's velocity $v_{x,\rm g}$ is offset and kinematically de-correlated from the stellar content and the bulk of the dark matter content $v_{x,\rm sdm}$. In fact, this gas component is homogeneous across $R_{\rm vir}^{\rm sat}$ and clearly static in the frame of the cluster. It is therefore composed of ICM fly-through particles rather than satellite bound gas. As the dark matter content is more diffuse, it is more accurate to compare the offset to where the stellar content is, at the heart of the halo (the actual galaxy). $B_g$ thus quantifies the relative velocity displacement of the gas content from the galaxy and can therefore be used as a more local and adaptive measure of the effect of stripping and strangulation on a given halo. 

\begin{figure*}
	\includegraphics[width=\textwidth]{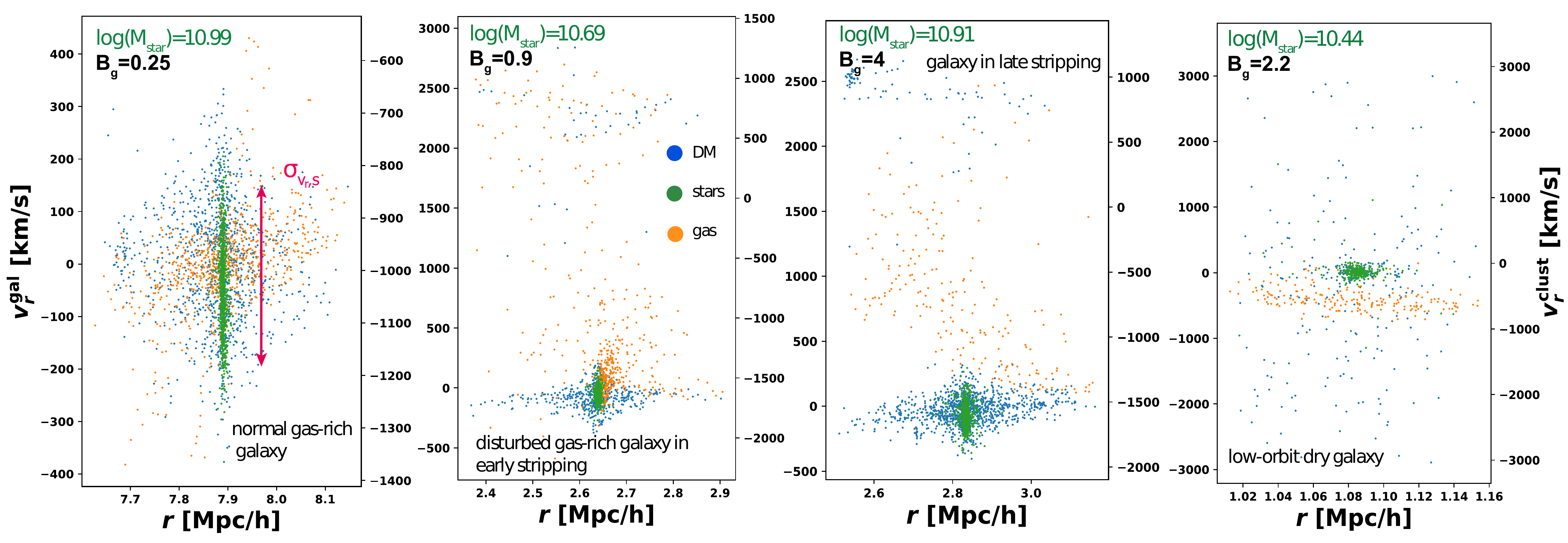}
    \caption{1-D phase-space diagrams for four different halos. Dark matter particles are in blue, stars in green, and gas in orange. $r$ is the cluster-centric position, velocities are in the galaxy frame on the left and in the cluster frame on the right. $v_r^{\rm clust}<0$ for infalling halos. $B_g$ is lower for bound, gas-rich situations (\textbf{two left panels}) and higher for unbound, gas-poor situations (\textbf{two right panels}).}
    \label{fig:gasbindscenarios}
\end{figure*}

Fig.~\ref{fig:gasbindscenarios} displays 1-D phase-space diagrams of selected halos to illustrate the evolution of $B_{\rm g}$ on a few examples. Note that these diagrams use the cluster-centric distance $r$ and radial velocities as the direction of analysis, to visually emphasize stripping.

The leftmost panel shows a typical gas-rich galaxy. Here, the dark matter, gas, and stellar content are all centred together in phase space, and the stellar content is distinctly tightly positioned at the centre of the halo, where velocity dispersion is high. The corresponding $B_{\rm g}$ is well below 1.The centre-left panel shows a gas-rich galaxy early in the process of stripping. Here, gas particles visibly vary in velocity, with a significant fraction reaching or exceeding the escape velocity, compared to the bulk of the gas, stars, and dark matter. $B_{\rm g}$ is close to 1. Depicted in the centre-right panel is a galaxy late in the process of stripping. There is little gas content associated with the same phase-space location as the dark matter and stellar content, with the majority of gas particles spanning a velocity tail that extends from the velocity of the halo's outskirts to the velocity of the ICM. One can see that the gas directly facing the centre of the cluster is escaping while the gas further away is still marginally synchronous with the galaxy. $B_{\rm g}$ is well above 2, boosted by the large differential velocity between the satellite halo and the cluster medium. The rightmost panel shows the same galaxy as Fig.~\ref{fig:gasbindexplan} whose gas content is fully disassociated kinematically from the stellar content, meaning the gas displayed here is purely the ICM, and the galaxy is dry and unable to form stars. Note that this galaxy displays almost no relative radial velocity towards the cluster, and is on a nearly pure tangential motion, nonetheless $B_{\rm g}$ is clearly above 2.

It should be noted that, albeit unfrequent, halos that do not contain gas particles at all (irrespective of temperature) due to the limited resolution of the ICM do not have an assigned $B_{\rm g}$ value. They are thus excluded from this analysis. It's however easy to combine both halos devoid of gas, $n(M_{\rm gas}=0)$, and halos with $B_{\rm g}>2$, $n(B_{\rm g}>2)$, to define the fraction of effectively dry halos:
\begin{equation}
f_{\rm dry}=\frac{n(B_{\rm g}>2)+n(M_{\rm gas}=0)}{N},
\label{eq:dryfrac}
\end{equation}
 in a sample of $N$ halos.
 
\section{Quenching of galaxies near and far from intra-cluster filaments}
\label{sec:quenching}
In this section, we use the galaxy properties outlined in ~\ref{sec:quetrac} (sSFR, $g-r$, $f_{\rm gas}^{\rm cold}$) to quantify the specific influence of intra-cluster filaments on the quenching of cluster galaxies. We discuss the mass-dependence of this effect in ~\ref{sec:masseff} and suggest possible origins of the effect in ~\ref{sec:contrast}.

\subsection{Survival of star forming galaxies near filaments}
\label{sec:sfg}

As a first step, we analyse how filaments locally influence the fraction of star forming galaxies in clusters. Recall that we define a star-forming galaxy as a galaxy with $sSFR>10^{-11}$ yr$^{-1}$, a cut chosen in accordance with observational papers  \citep{Wetzel_2013, roberts}.

\begin{figure}
	\includegraphics[width=\columnwidth]{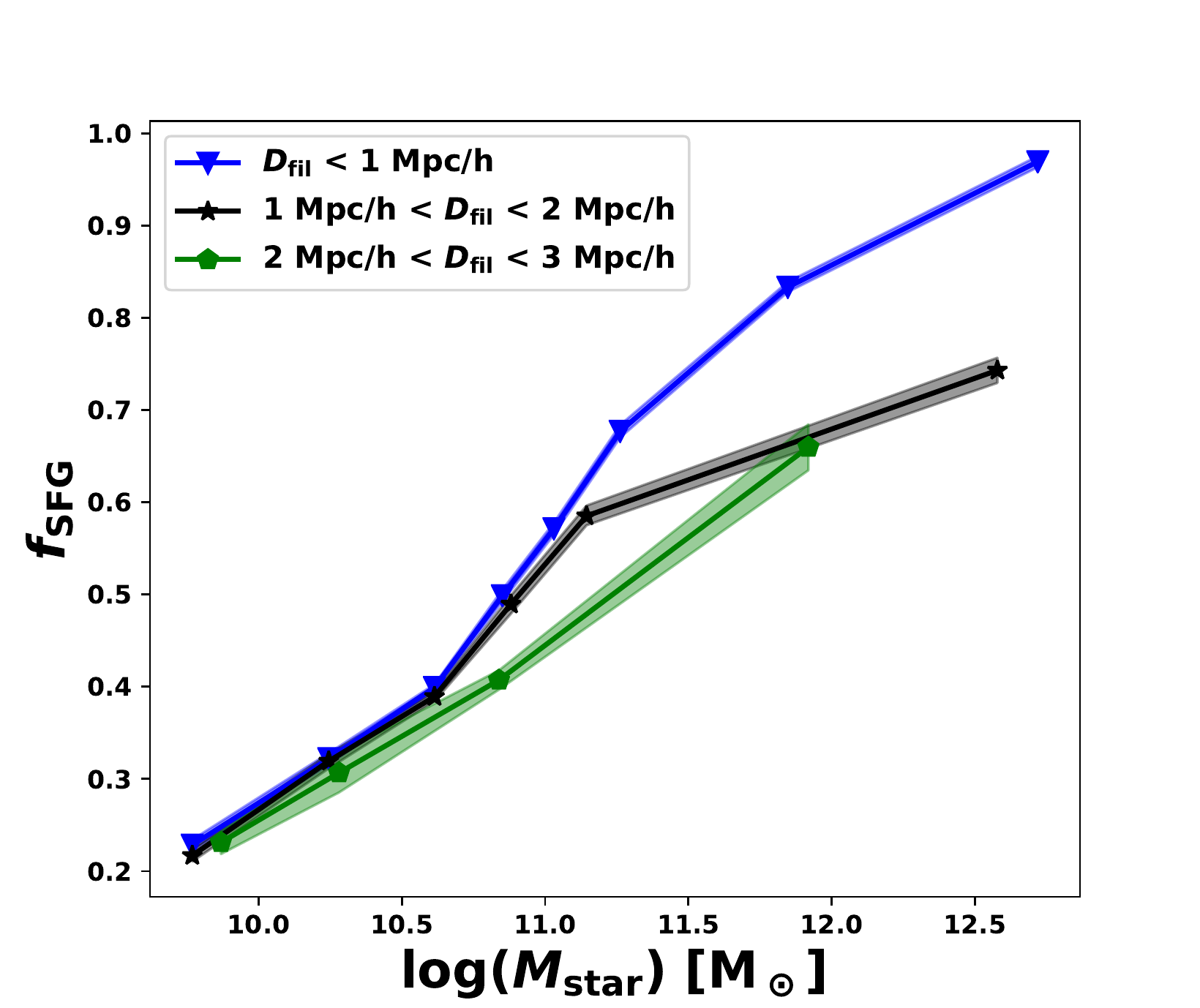}
    \caption{Evolution of the fraction of star forming galaxies, $f_{\rm SFG}$, with stellar mass, split by filament distance $D_{\rm fil}$, for all halos within 1 $R_{\rm vir}$ of their cluster. At fixed mass, the closer to a filament, the higher the fraction of star forming galaxies. Shaded contours are 1-$\sigma$ bootstrap errors.}
    \label{fig:fSFG}
\end{figure}

In Fig.~\ref{fig:fSFG}, galaxies within one virial radius of their host cluster ($R_{\rm vir}$) are split into different bins based on their distance to the nearest filament, $D_{fil}$, and the fraction of star forming galaxies per stellar mass bin is presented versus the stellar mass. Shaded contours display 1-$\sigma$ bootstrap errors, with each stellar mass bin containing at least 750 halos. We do not investigate beyond 3 $h^{-1}$.Mpc as there are very few galaxies this far from a filament in cluster environments.

Irrespective of distance to filament, the fraction of star forming galaxies steadily increases with stellar mass across the whole range investigated. At low masses, this trend aligns with established work in that the specific star formation rate of galaxies increases steadily with stellar mass. However, as galaxies grow and quench, this relationship is expected to reverse, which in this simulation is limited to a flattening at high mass as we discussed in Section~\ref{sec:sfr} due to numerical artifacts detailed in~\ref{sec:sfr}. For this reason, we limit our analysis to  comparisons at fixed stellar mass. Additionally, the kink observed at $10^{10.5} M_{\odot}$ on Fig.~\ref{fig:fSFG} is also numerical in origin and corresponds to the mass at which AGN feedback becomes active in the simulation.

At fixed stellar mass, Fig.~\ref{fig:fSFG} shows that galaxies are more star forming closer to filaments. Below $10^{10.5} M_{\odot}$ this trend is visible for galaxies less than 2 $h^{-1}$.Mpc from a filament (green to black). It becomes increasingly significant for galaxies within 1 $h^{-1}$.Mpc (green/black to blue) at stellar masses above $10^{11} M_{\odot}$. This result suggests that filament proximity increases the ability of cluster galaxies to form stars, irrespective of stellar mass. 

We investigate this trend in the next section using SDSS galaxy colours, which do not suffer the same limitations as sSFR.

\subsection{Galaxy colours around intra-cluster filaments}
\label{sec:galcolsan}

In the following, we show that the trend of filaments being a region of increased star formation is recovered with higher significance for standard SDSS galaxy colours. 

\begin{figure*}
	\includegraphics[width=1.9\columnwidth]{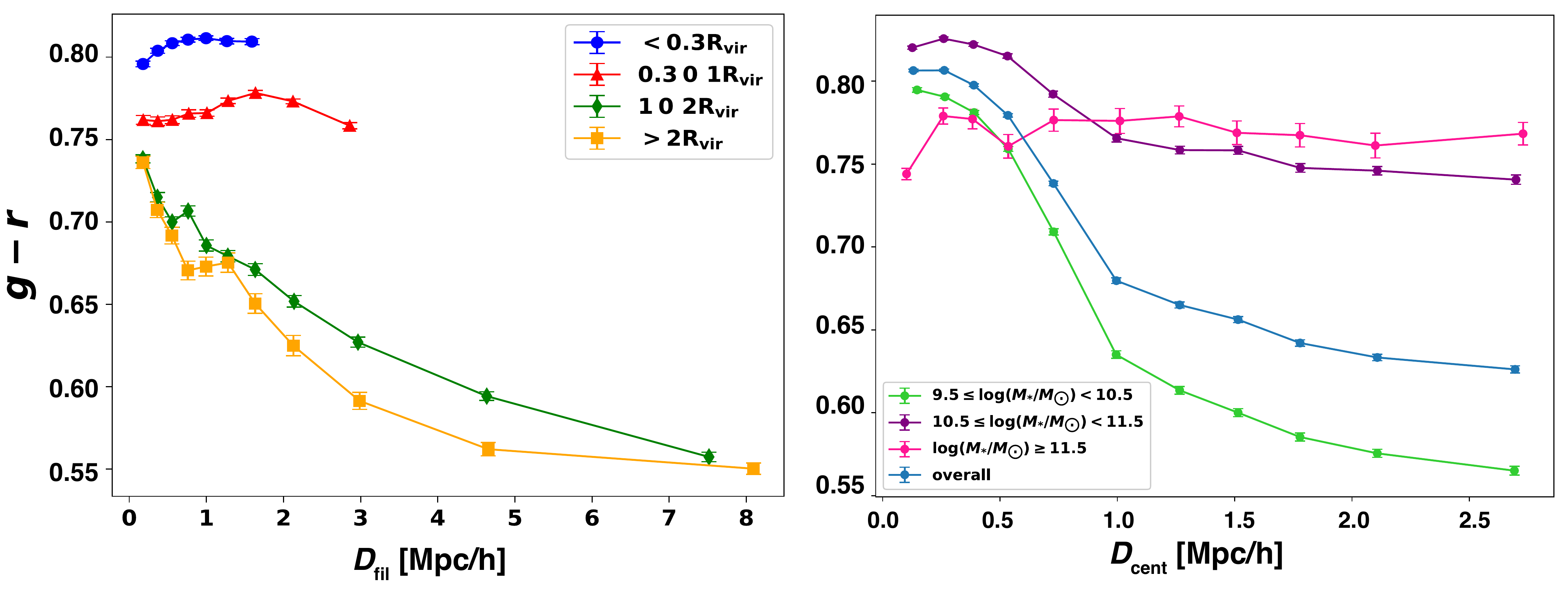}
    \caption{\textbf{Left Panel:} Mean colour $g-r$ versus distance to filament $D_{\rm fil}$, split by cluster-centric distance. Galaxies outside $R_{\rm vir}$ are redder close to field filaments. In contrast, galaxies inside $R_{\rm vir}$ exhibit a decrease in redness close to intra-cluster filaments. \textbf{Right Panel:} Mean $g-r$ versus distance to centre of cluster $D_{\rm cent}$, for four different stellar mass bins. The majority of galaxies redden towards the cluster centre, except for massive group centrals. All error bars are standard error on the mean.}
    \label{fig:colcombo}
\end{figure*}

In the left panel of Fig.~\ref{fig:colcombo} galaxies are split by cluster-centric distance and their mean $g-r$ colours plotted against their distance to the nearest filament. Error bars are standard error on the mean.

At fixed distance to filament, $D_{\rm fil}$, galaxies are redder at smaller cluster-centric distances. Galaxies residing at 1-2 $R_{\rm vir}$  and >2 $R_{\rm vir}$ are roughly similar with respect to $g-r$ colours at fixed $D_{\rm fil}$, and are noticeably bluer than galaxies residing in the cluster, at <0.3 and 0.3-1 $R_{\rm vir}$. This difference clearly demarcates the expected environmental differences inside and outside the cluster, where galaxies inside a cluster's virial radius are significantly redder due to quenching.

Galaxies further than 0.3 $R_{\rm vir}$ from cluster centre are progressively redder closer to filaments up until 2 $h^{-1}$.Mpc away, an increase of approximately 0.12 in all cases. Galaxies at 1-2 $R_{\rm vir}$ and >2 $R_{\rm vir}$ continue to redden even closer to filaments. This trend beyond the cluster virial radius is consistent with what is observed for filaments in the field, which host on average more massive galaxies than their surroundings (Appendix~\ref{sec:mstevolapp}) and are increasingly considered a region of pre-processing, due to increased turbulence and interactions \citep{kraljic, Laigle_2017, Sarron_2019}. On the contrary, those galaxies in the inner two cluster-centric bins display a smooth decrease in redness of approximately 0.02 coming within 2 $h^{-1}$.Mpc from a filament, despite the fact that they are still on average more massive than their counterparts further away (Appendix~\ref{sec:mstevolapp}). This novel trend inside the cluster indicates that intra-cluster filaments tend to alleviate the reddening of galaxies in their vicinity.

While this effect may seem small compared to a seemingly large increase of $g-r$ towards filaments at $D_{\rm fil}>2$ $h^{-1}$.Mpc it is important to remember that distance to filament, $D_{\rm fil}$, and distance to the centre of the cluster, $D_{\rm cent}$, are correlated and therefore the specific effect of filaments must be disentangled from the increase of $g-r$ with proximity to the cluster's core. 
Indeed, by construction galaxies closer to the cluster centre have smaller distances to filament on average. Most of the $g-r$ increase towards the filaments at $D_{\rm fil}>2$ $h^{-1}$.Mpc, observed within the cluster on the left panel, can actually be traced back to galaxies being also closer to the node. One should also note that galaxies closest to filaments also tend to be more massive and therefore redder. It is a mass-driven effect, not a genuine correlation with the filament.

On the right panel of Fig.~\ref{fig:colcombo}, mean $g-r$ galaxy colour is plotted against cluster-centric distance $D_{\rm cent}$, separated into stellar mass bins. The trend for all galaxies up to $10^{11.5} M_{\odot}$ indicates that galaxies indeed redden ($g-r$ increases) as they approach the cluster centre. This increase is most marked inside the cluster virial radius, consistent with established expectations for galaxies, which quench as they fall into clusters \citep{balogh_2000, haines, Raichoor_2012, Pintos_Castro_2019}. 

Beyond the virial radius ($D_{\rm cent}\geq1 R_{\rm vir}$), the most massive galaxies are the reddest, and the least massive galaxies are bluest, with a $g-r$ separation of at least 0.15. This trend is consistent with observations of galaxies in the field \citep{Strateva_2001, Baldry_2004}. 

However, note that galaxies in all mass bins display a difference in $g-r$ colour of less than 0.08 from 1 to 2 $R_{\rm vir}$ from the cluster. This indicates that the proximity to the cluster has little effect on galaxy colour beyond the cluster virial radius, consistent with more limited quenching for galaxies residing at 1-2 and >2 $R_{\rm vir}$ shown in the left panel. 

The rarest, most massive ($M_{\rm star}>10^{11.5} M_{\odot}$) galaxies are largely unchanged in $g-r$ colour irrespective of their cluster-centric distance. This is due to the fact that most of these galaxies are mostly centrals of their own massive groups, and are therefore less impacted by the main cluster than by their local halo environment. 

 To summarize, evolution of $g-r$ with $D_{\rm cent}$ indicates that the galaxies closest to the cluster centre are the reddest, and that the more massive galaxies are also the reddest, excluding centrals of their own massive groups, secondary nodes of the cosmic web, which are largely sub-dominant outside the inner 0.3 $R_{\rm vir}$ of clusters and virtually absent outside the inner 0.3 $h^{-1}$.Mpc from filaments (innermost $D_{\rm fil}$ bin on the left panel). As such, the population of cluster galaxies less than 2 $h^{-1}$.Mpc from filaments is dominated by these galaxies reddening with mass, and thus while the decrease in redness of cluster galaxies close to filaments exhibited may seem small, it is actually significant. We stress again that, since this decrease in redness cannot be attributed to higher mass or proximity to the  cluster centre (which would produce the opposite trend), it is in fact an effect of the filament environment.

The fact that the colour becomes bluer and thus the star formation exhibits this local upturn due to intra-cluster filaments suggests that there must be an uptick in the amount of cold gas available to galaxies in the vicinity of filaments, which we explore in the following section.

\subsection{Cold Gas Fraction}
\label{sec:gasfracan}
Galaxies quench when they are unable to refresh or retain their cold gas reserves. In the following we analyse how intra-cluster filaments affect the cold gas stores of cluster galaxies in their vicinity, by way of the cold gas fraction (as defined by Equation~\ref{eq:CGF}) of halos.

\begin{figure*}
	\includegraphics[width=\textwidth]{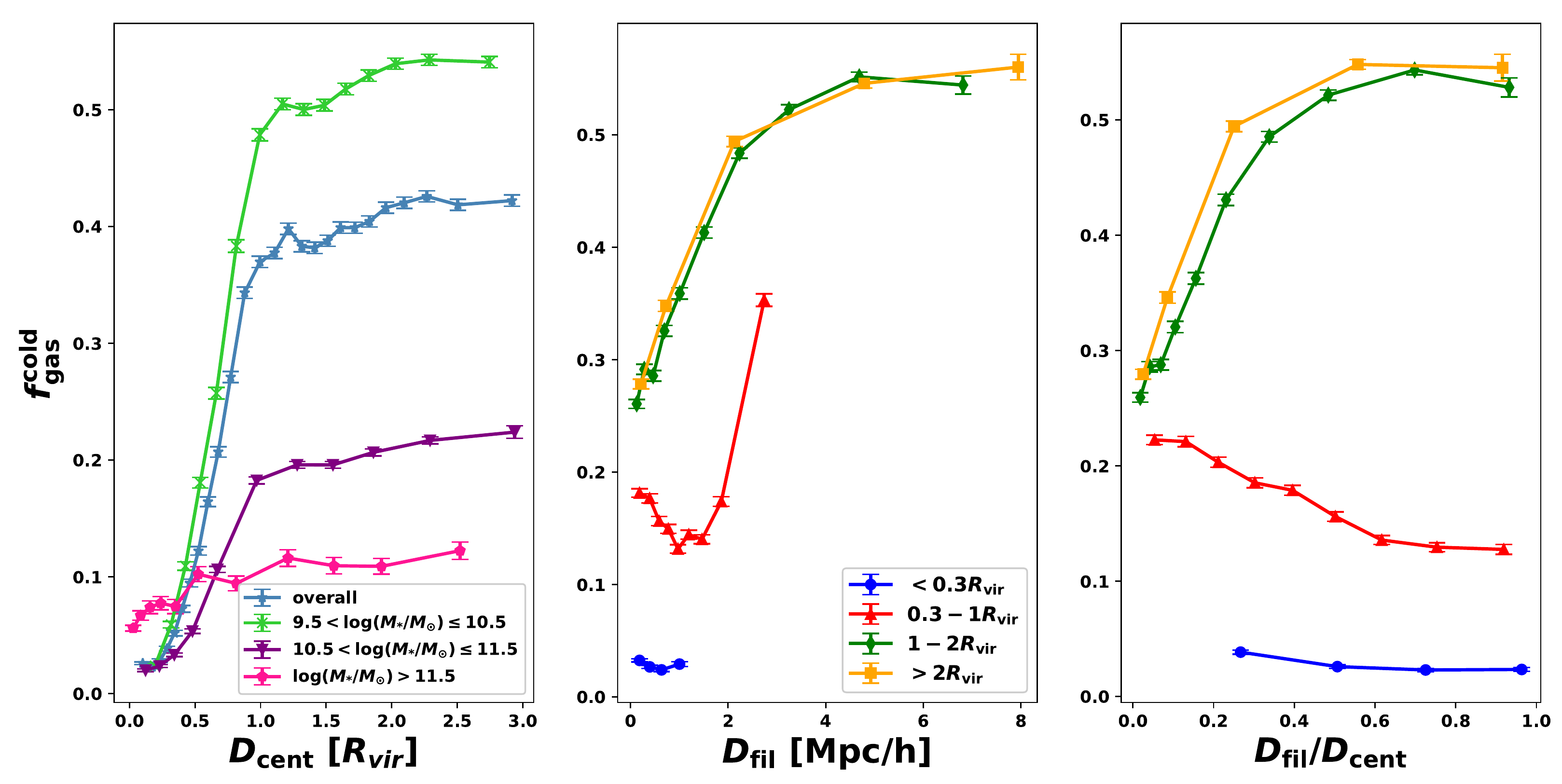}
    \caption{\textbf{Left Panel:} Mean cold gas fraction ($f_{gas}^{cold}$) versus distance to centre of cluster, for different stellar mass bins. Halos lose cold gas as they approach the cluster centre. \textbf{Centre Panel:} Mean $f_{gas}^{cold}$ versus distance to filament, split by cluster-centric distance. Halos outside $R_{\rm vir}$ progressively lose their cold gas closer to filaments. Comparatively, halos inside the $R_{\rm vir}$ exhibit an uptick in cold gas closer to filaments. \textbf{Right Panel:} Mean $f_{gas}^{cold}$ versus angular separation ($D_{\rm fil}/D_{\rm cent}$). Within clusters, $f_{gas}^{cold}$ increases at lower angular separations. All error bars are standard error on the mean, with $\approx 5000$ halos per bin.}
    \label{fig:cgfcombo}
\end{figure*}

The left panel of Fig.~\ref{fig:cgfcombo} shows the evolution of the mean cold gas fraction of halos with their cluster-centric distance, split by stellar mass. Error bars indicate standard errors on the mean, with 5000 halos per bin, except for the curve with the most massive halos with 300 halos per bin. 

Overall (light blue), the cold gas fraction of halos decreases approaching the cluster centre. This decrease is minimal in the cluster outskirts, then sharp  inside 1 $R_{\rm vir}$. This is consistent with the result obtained for galaxy colours in the previous section and simply indicative of expected environmental quenching.

The cold gas fraction of halos decreases by at most 0.05 from 3 to 1 $R_{\rm vir}$ in any given mass bin. Galaxies below $10^{10.5} M_{\odot}$ (light green) exhibit the largest drop in cold gas fraction, from approximately 0.5 to approximately 0.03, while more massive galaxies (purple) drop from approximately 0.2 to approximately 0.03. This is aligns with the expectation that galaxies are stripped of their cold gas stores by ram pressure as they fall through the intra-cluster medium \citep{Wetzel_2013, roberts, maier2019slow}, with the smaller galaxies having a weaker gravitational potential to hold onto their cold gas. This figure again demonstrates that galaxy quenching is strong and progressive within the simulated clusters.

The most massive galaxies (above $10^{11.5} M_{\odot}$, in pink) consistently maintain a cold gas fraction of approximately 0.1, dropping by approximately $40\%$ only in the heart of clusters. This is again in line with the fact that these are most often the centrals of massive groups or small clusters, affected by their local environment rather than the main cluster environment.

We examine the impact of cosmic filaments in the middle panel of Fig.~\ref{fig:cgfcombo}, which shows the mean cold gas fraction of halos in relation to their distance to closest filament, again split by cluster-centric distance. Error bars indicate standard errors on the mean, with 5000 halos per bin. 

Halos residing  within 1-2 $R_{\rm vir}$ and at distances >2 $R_{\rm vir}$ away from the cluster centre have similar average cold gas fractions at fixed distance to filament $D_{\rm fil}$, but this gas fraction decreases sharply with decreasing $D_{\rm fil}$.

This is consistent with the fact that filament galaxies are on average more massive hence redder and more gas-poor than their counterparts further away \citep{kraljic, welker}.

Moreover, field filaments are much denser and the site of markedly increased galaxy numbers than their surrounding voids and walls, where galaxies are more isolated and evolve in a lower density, less violent environment. This increase in galaxy count near filaments translates into an increase in galaxy interactions and mergers. Recent studies suggest that this increase in interactions and turbulence drives some early quenching or pre-processing in filament galaxies \citep{kraljic, Laigle_2017, Sarron_2019}. This likely contributes to the steep decrease of $f_{\rm gas}^{\rm cold}$ towards filaments for galaxies outside $R_{\rm vir}$, as is investigated in the next subsection.

A different trend is observed around filaments between 0.3-1 $R_{\rm vir}$. The mean cold gas fraction increases by approximately 50$\%$ towards the spine  across the inner 1 $h^{-1}$.Mpc to an intra-cluster filament. This upturn is in line with our results on galaxy colour, again indicating that intra-cluster filaments are effectively delaying quenching. 

Note however that, at fixed $D_{\rm fil}$, the mean cold gas fraction for halos beyond $R_{\rm vir}$ is always higher by $10\%$ or more  than halos in the 0.3-1 $R_{\rm vir}$ bin. This is expected as halos transition from the field environment to the dense, hot cluster environment where cold gas is heated up and$/$or stripped, as per the left panel, and this effect remains dominant in predicting the cold gas fraction of a satellite halo.

To accommodate for the bias whereby halos residing closer to the cluster centre are by definition closer to a filament, we repeat our analysis with $D_{\rm fil}/D_{\rm cent}$, the angular separation of a halo/galaxy to its nearest filament, in the right panel of Fig.~\ref{fig:cgfcombo}. We see that the halos outside the cluster virial radius still exhibit the same strong decrease in cold gas fraction at small filament separations, consistent with the centre panel and with the fact that $f_{\rm gas}^{\rm cold}$ shows little variation with $D_{\rm cent}$ outside $R_{\rm vir}$.

But this new measure reveals a strong, progressive increase of the average cold gas fraction as the angular separation to an intra-cluster filament decreases (red). This occurs despite the increase in stellar mass associated with filament proximity and clearly points towards the ability of cluster galaxies residing near filaments to retain or accrete cold gas. 
In the next section, we proceed to specifically investigate further the stellar mass dependence of the trends described in this and the previous sections.

\subsection{Stellar mass dependence.}
\label{sec:masseff}
We have established that halos hosting galaxies appear bluer and contain more cold gas close to intra-cluster filaments than their counterparts further away, while halos outside clusters display the opposite trend. It is important to ensure that these effects are due to filaments and not merely due to a halo's stellar mass as there is a strong increase in stellar mass near the spine of filaments (Appendix~\ref{sec:mstevolapp}). More generally, the mass dependence of such trends is interesting to analyse. We therefore analyse the direct relationship of the cold gas fraction with stellar mass in this section.

\begin{figure}
	\includegraphics[width=\columnwidth]{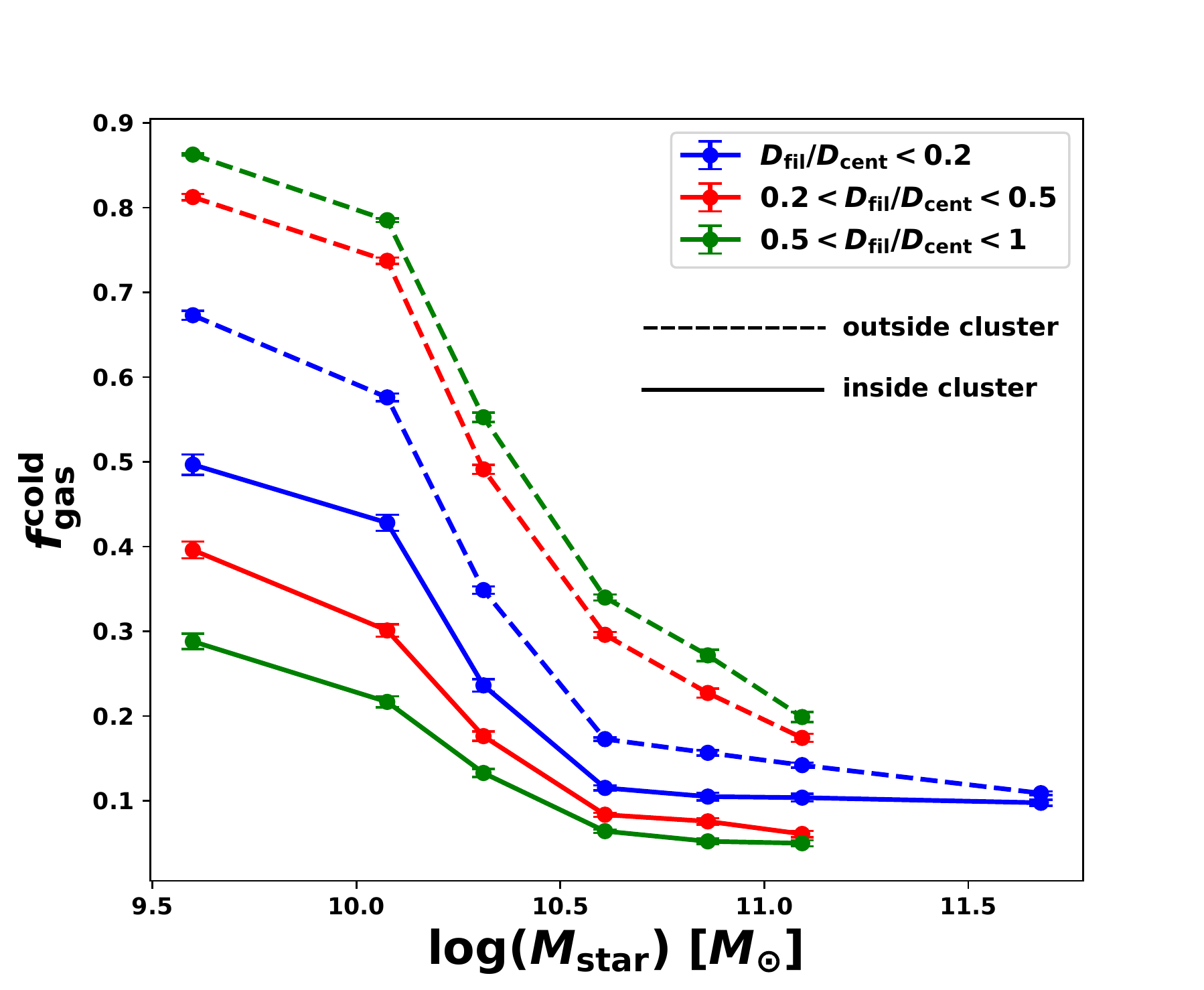}
    \caption{Average cold gas fraction ($f_{\rm gas}^{\rm cold}$) versus stellar mass, binned by angular separation to filament and residency inside or outside a cluster. $f_{\rm gas}^{\rm cold}$ decreases with increasing stellar mass. Outside clusters, $f_{\rm gas}^{\rm cold}$ decreases at smaller angular separations. Inside a cluster, it conversely increases as angular separation decreases.All errors are standard error on the mean. The smallest bin contains at least 350 halos.}
    \label{fig:cgfcontrol}
\end{figure}

Fig.~\ref{fig:cgfcontrol} shows the average cold gas fraction for halos against their stellar mass, distinguished by both angular separation $D_{\rm fil}/D_{\rm cent}$ (colours) and residency inside or outside the cluster virial radius (dashed vs. solid lines).

Across all angular separations and distances to the cluster centre, the average cold gas fraction markedly decreases with increasing stellar mass. This is expected as galaxies use up their cold gas stores to form stars as they become more massive. It also confirms that the increase in cold gas fraction exhibited by halos near intra-cluster filaments in Fig.~\ref{fig:cgfcombo} is not correlated to the increase in stellar mass close to filaments.

The most massive halos, with stellar masses greater than $10^{11.5} M_{\odot}$, are found exclusively along the spine of cosmic filaments (often as secondary nodes of the cosmic web), hence reside at the smallest angular separations (blue). Noticeably, there is less than 0.02 difference in cold gas fraction between such massive halos outside (dashed) versus inside (solid) the cluster environment, the smallest difference across any fixed mass. This re-iterates that these halos are centrals of massive groups which do not feel the effects of the transition to, and journey through, the cluster environment.

At fixed stellar mass outside the cluster environment (dashed lines), cold gas fraction decreases as angular separation to filament ($D_{\rm fil}/D_{\rm cent}$) decreases. This effect is present irrespective of stellar mass and largest for the lowest mass halos. This is in line with our previous results but now indicates clear pre-processing of field galaxies in filaments outside clusters at all stellar masses.

Opposing this, at fixed stellar mass inside the cluster environment (solid lines), cold gas fraction increases as angular separation to filament decreases, again most effectively for the lowest mass halos, although apparent across the stellar mass range. This indicates that the halting of quenching we established in these regions is directly caused by the presence of the intra-cluster filaments and efficient at all masses, rather than an effect of the mass distribution. We discuss this novel contrast in the following section.

\subsection{DM Halo mass dependence.}
\label{sec:masseffh}

Estimating the DM halo mass dependence of such trends is crucial. We verified that DM mass increases towards filaments in a fashion similar to the stellar mass. However, a given stellar mass can correspond to widely varying DM halo masses depending on whether we are considering the central galaxy of a satellite halo (a {\it genuine satellite}) or a smaller mass satellite in a larger DM satellite halo, i.e a``satellite of another satellite", which we refer to as a {\it sub-satellite}. Genuine satellites evolve under the direct influence of filaments and their host cluster while sub-satellite can also undergo pre-processing within their primary host.

For simplicity, we therefore limit ourselves to genuine satellites in the following section. Sub-satellites' evolution is qualitatively similar and secondary variations are out of the scope of the present study. For a more detailed analysis of how filaments impact pre-processing inside clusters, we refer the reader to our upcoming follow-up paper (Welker et al. 2021, in prep.)

\begin{figure}
	\includegraphics[width=\columnwidth]{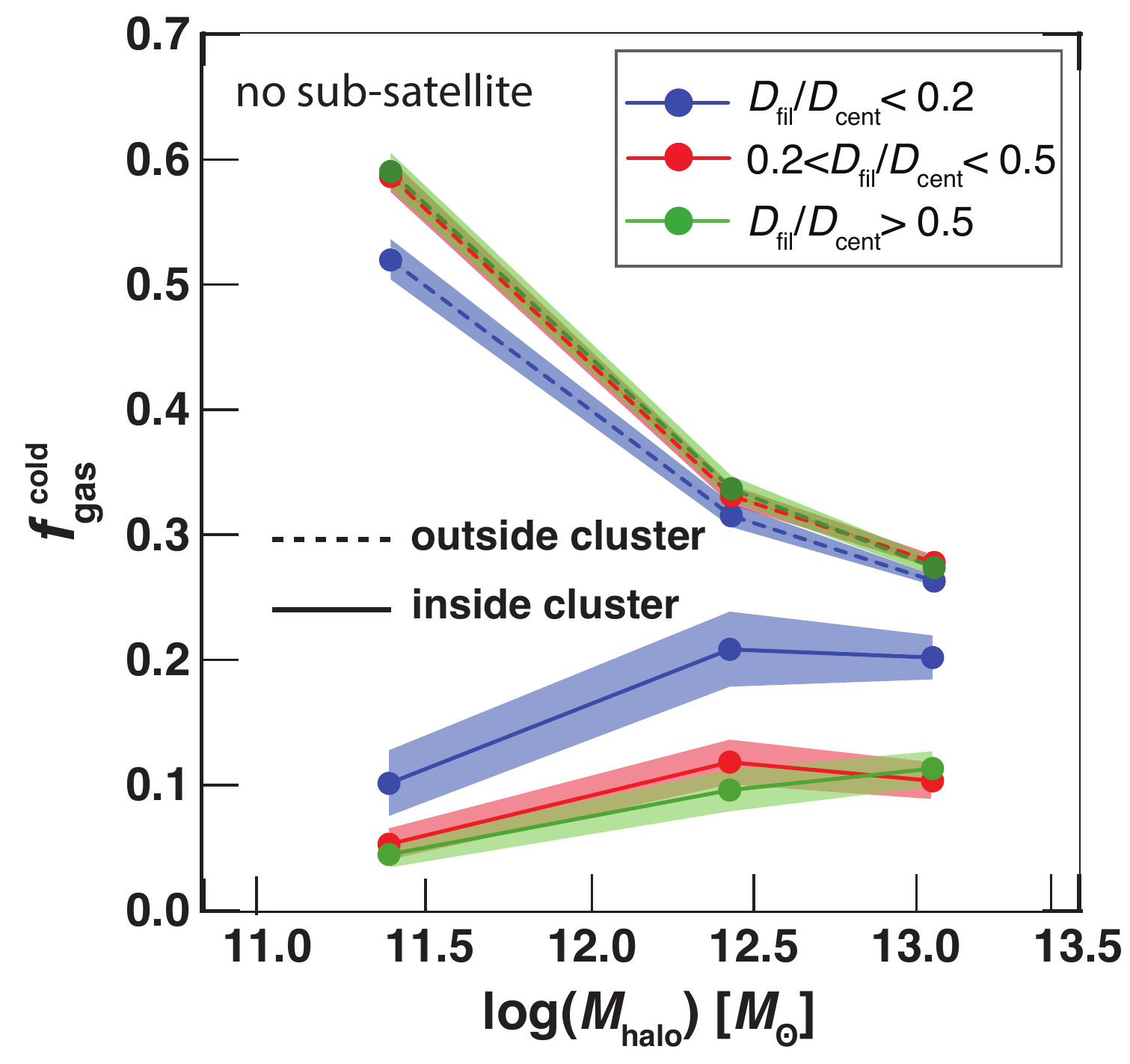}
    \caption{Average cold gas fraction ($f_{\rm gas}^{\rm cold}$) versus DM halo mass for all genuine satellites, binned by angular separation to filament and residency inside or outside a cluster. Outside clusters, $f_{\rm gas}^{\rm cold}$ decreases at smaller angular separations. Inside a cluster, it conversely increases as angular separation decreases.}
    \label{fig:cgfcontrolh}
\end{figure}

In Fig.~\ref{fig:cgfcontrolh}, we reproduce the analysis of Fig.~\ref{fig:cgfcontrol} replacing stellar mass by DM halo mass for all genuine satellites. It shows the average cold gas fraction for halos against their DM halo mass, distinguished by both angular separation $D_{\rm fil}/D_{\rm cent}$ (colours) and residency inside or outside the cluster virial radius (dashed vs. solid lines). Shaded areas are standard error on the mean. 

At fixed halo mass, the cold gas fraction of cluster satellites decreases away from the filament, showing once again that this behaviour is not just a tracer of more massive halos lying closer to filaments. The opposite trend is observed outside the cluster, once again consistent with increased density-related quenching in filaments lying outside clusters.

Interestingly, this analysis also confirms that, inside the cluster, the cold gas fraction decreases at lower DM halo mass, consistently with low-mass halos being more easily stripped and quenched than their massive counterparts. This trend was however lost with stellar mass as:
\begin{itemize}
    \item higher stellar masses also reflect increased conversion of cold gas into stars.
    \item low-stellar mass bins on Fig.~\ref{fig:cgfcontrol} include galaxies that are sub-satellites, hence better protected from accretion shocks at entry into the cluster (see Welker et al. 2021, in prep)
\end{itemize}

\subsection{Contrasting intra-cluster filaments with their field counterparts}
\label{sec:contrast}
The fact that the impact of intra-cluster filaments on local galaxies run contrary to those of field filaments may seem surprising. It should however be stressed that in the field, filaments are a region of increased interaction and disruption compared to the lower density background governed by less turbulent cosmic flows \citep{AragonCalvo2019}. However, in clusters, interactions and stripping by the dense, hot, turbulent intra-cluster medium are ubiquitous and dominant effects. Thus, while field filaments represent a more dynamic environment relative to their surroundings, intra-cluster filaments may on the contrary be quiet in comparison to the rest of the cluster. Indeed, the ability of filaments to channel colder, more laminar gas streams along their spine deep into the cluster, where they might later diffuse locally, may help galaxies resist quenching from ram pressure stripping or strangulation, either by allowing more gas accretion onto galaxies or by lowering the local ram pressure onto halos. 

\begin{figure}
	\includegraphics[width=\columnwidth]{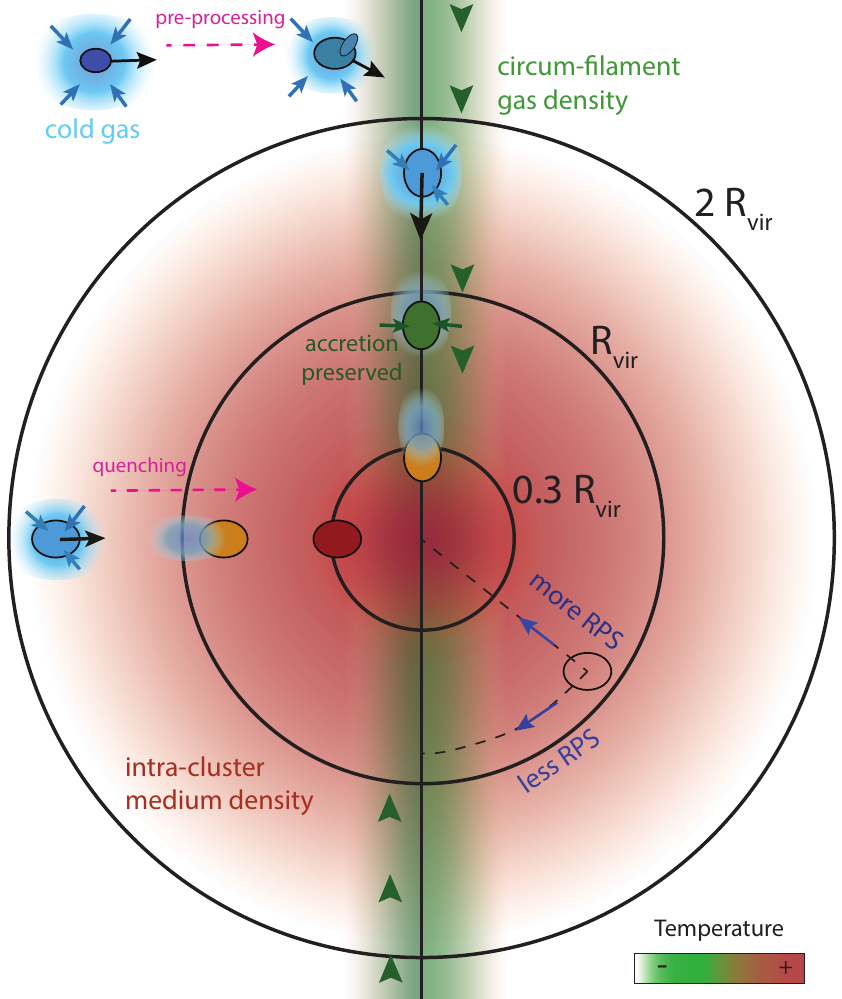}
    \caption{Illustration of observed trends and proposed mechanisms. Outside the cluster, galaxies are pre-processed near dense filaments. Inside clusters,  galaxies closer to filaments experience less quenching (hence remain bluer) due to reduced strangulation and ram pressure stripping (RPS) in the cooler, more coherent filament flows.}
    \label{fig:sketch}
\end{figure}

Fig.~\ref{fig:sketch} summarizes these expectations and illustrates this scenario, which we investigate in Section~\ref{sec:dynamics} by carefully analyzing the dynamics of both the intra-cluster medium and galaxies around filaments.

\section{Halting quenching: dynamics of gas and galaxies around filaments}
\label{sec:dynamics}
Having established that cluster galaxies in proximity to cosmic filaments exhibit telltale signs of delayed quenching, we investigate the dynamics of halos and gas flows around filaments in the cluster environment to understand the origin of this effect.

\subsection{Density and temperature of intra-cluster filaments}
\label{sec:filprofs}
As gas (cold gas in particular) is crucial to star formation, we examine the gas around filaments to understand the intra-filament medium and how it delays the quenching of cluster galaxies.

\begin{figure}
	\includegraphics[width=\columnwidth]{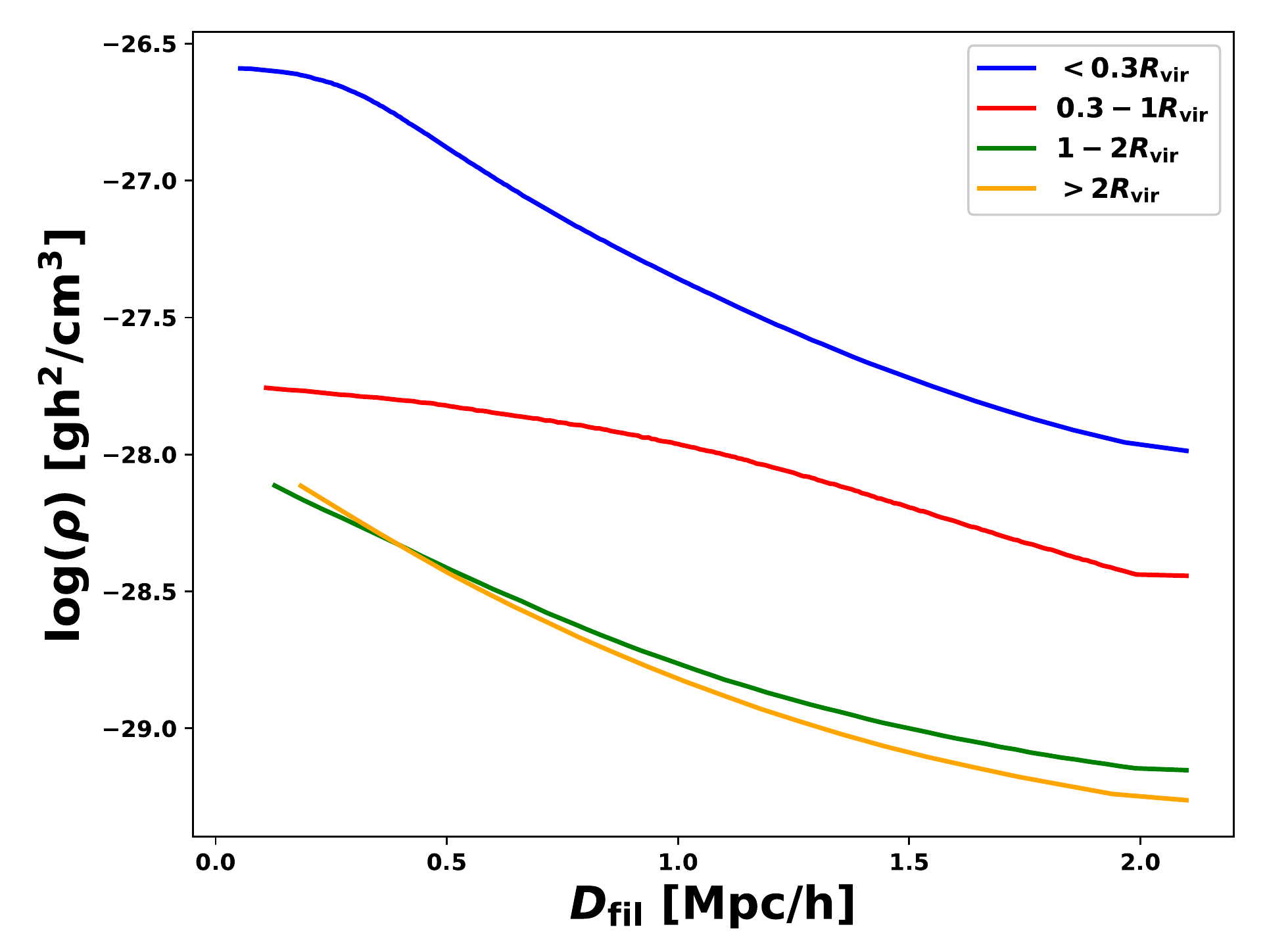}
    \caption{Evolution of the ICM diffuse gas density with distance to filaments.  Density increases towards the spine of filaments but filaments are little contrasted within $R_{\rm vir}$. All errors on the mean are below the 0.01 dex level and are therefore omitted.}
    \label{fig:densprof}
\end{figure}

Fig.~\ref{fig:densprof} shows the cylindrical density profile of diffuse gas around filaments in our simulation. Gas is binned by its cluster-centric distance, and we exclude all gas within the virial radius $R_{\rm vir}^{\rm sat}$ of satellite halos in order to focus on the background gas profile of the intra-cluster medium rather than the gas already within halos. We verified however that including gas in halos did not qualitatively modify the results. We present volume-weighted average values, with no temperature cut.

Irrespective of distance to the centre of the cluster, all curves show a progressive increase in diffuse gas density approaching the spine of filaments. This is to be expected, as by definition density peaks at the spine of cosmic filaments. However, while the overall increase in density of diffuse gas from 2 $h^{-1}$.Mpc from a filament to its spine is at least a factor of 10 for the 1-2 and >2 $R_{\rm vir}$ bins, it is reduced to a factor of 3 in the 0.3-1 $R_{\rm vir}$ bin. This decrease in the density contrast of diffuse gas filaments within $R_{\rm vir}$ is even more striking when correcting for the correlation between $D_{\rm fil}$ and $D_{\rm cent}$ , which further reduces the filament contrast to less than a factor of 2 within $R_{\rm vir}$, without impacting the contrast outside $R_{\rm vir}$. Filaments are therefore not very contrasted in terms of density within the cluster.

In comparison, at fixed distance to filament ($D_{\rm fil}$), gas is markedly denser closer to cluster centre. Moving from outside the cluster (1-2 and >2 $R_{\rm vir}$) to the 0.3-1 $R_{\rm vir}$ region, an increase in density by at minimum a factor of 3 is present ($\approx 6$ on average), and similarly moving from that bin to the deepest cluster-centric bin (<0.3 $R_{\rm vir}$), where it can even reach more than 10. This shows that clusters are increasingly dense environments toward their centres and that this density increase is dominant over the filamentary increase within $R_{\rm vir}$.

Therefore, while density-induced quenching mechanisms are relevant for field filaments, we cannot expect that same impact from intra-cluster filaments. This is re-visited in Section~\ref{sec:rpsan} when we examine ram pressure stripping in detail.

Gas closest to cluster centre exhibits the largest increase in density approaching the filament spine. However, once again, for gas within 0.3 $R_{\rm vir}$ motion towards the closest filament is strongly correlated to motion towards the cluster centre where the central node of filaments resides. As such, at this depth within clusters, increasing cluster density is a more relevant factor than filaments. We checked that looking at evolution of density with $D_{\rm fil}/D_{\rm cent}$ rather than with $D_{\rm fil}$ did effectively reduce the maximal amplitude of variation across observed range from 1.4 dex to 0.6 dex for halos within 0.3 $R_{\rm vir}$, and from 0.5 dex to $<0.1$ dex for halos within 0.3-1 $R_{\rm vir}$.

While the density contrast of intra-cluster gaseous filaments is limited, this may not be the case for their temperature contrast.

Indeed, in the field, as it collapses into filaments, gas shocks and cools as it radiates energy and entropy away, so the spine of filaments may funnel cooler gas towards the nodes of the cosmic web \citep{Katz_1993, brooks2009, Keres_2009, kleiner}. 

However, whether cold flows persist deep into the cluster environment has been subject to debate as intense enough streaming instabilities, active galactic nuclei, and supernova feedback might shred them apart \citep{Powell_2011, Dubois13, Zinger_2016,Zinger_2018}.

Nonetheless, irrespective of what disturbs these high redshift streams in clusters, they might still pervade the ICM in the form of temperature anisotropies, detectable at $z=0$.

\begin{figure*}
	\includegraphics[width=\textwidth]{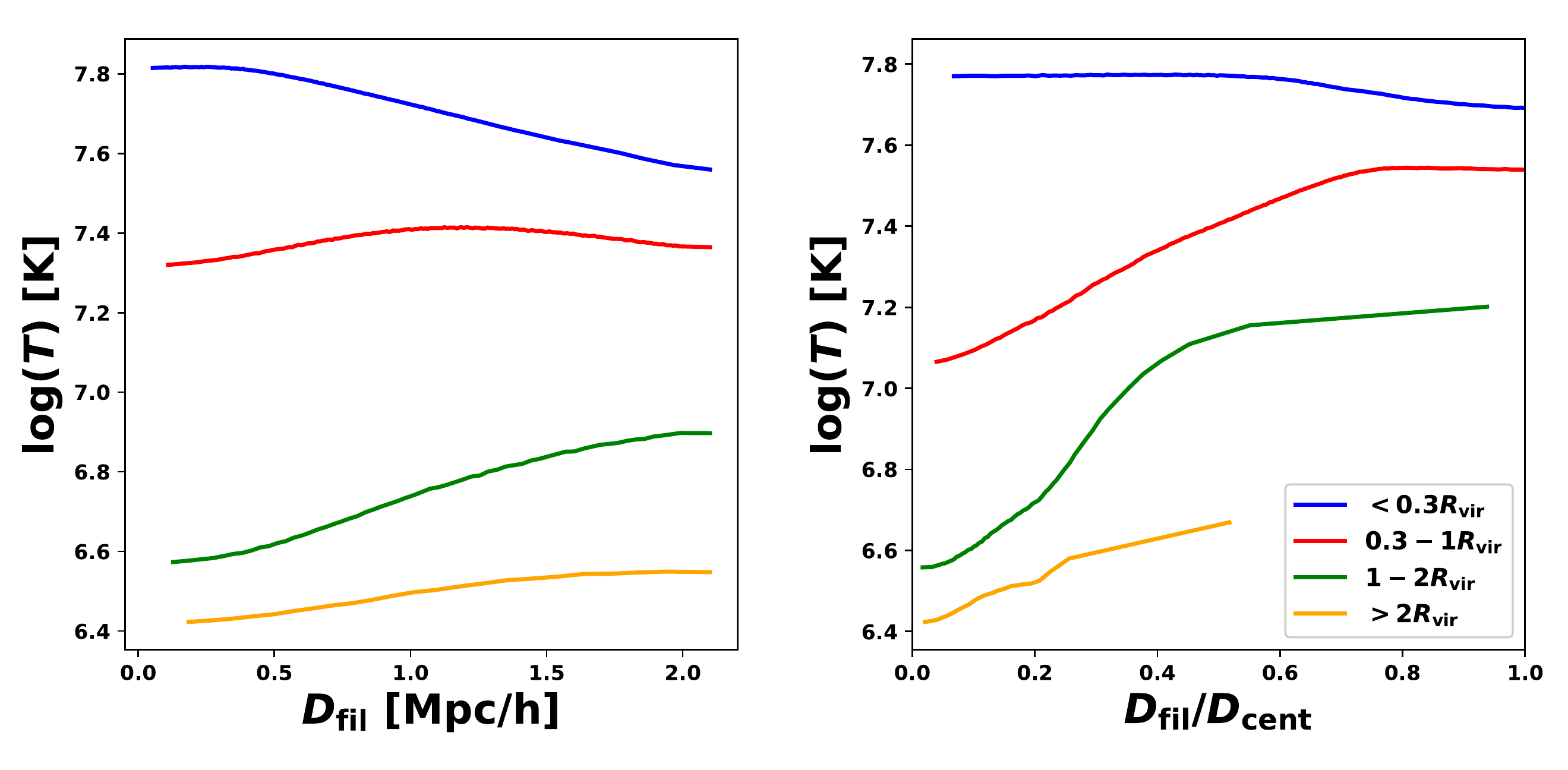}
    \caption{Mean temperature of the diffuse ICM gas versus distance to filament (\textbf{Left Panel}) and angular separation (\textbf{Right Panel}). Filaments persist as cooler gas streams down to 0.3 $R_{\rm vir}$. All errors on the mean are below the 0.01 dex level and are therefore omitted.}
    \label{fig:tempcombo}
\end{figure*}

Fig.~\ref{fig:tempcombo} shows the temperature profile of gas around filaments, cylindrically in the left panel and with angular separation ($D_{\rm fil}/D_{\rm cent}$) to filament in the right panel. Gas is again separated by cluster-centric distance, and gas within the virial radius of halos is excluded. Values shown are volume-weighted averages.

 Gas residing further than 0.3 $R_{\rm vir}$ from cluster centre does exhibit progressive cooling approaching filaments radially (left panel), an effect that is even sharper at decreasing angular separation (right panel). This demonstrates that filaments persist as cooler flows down to this depth in the cluster environment. Notably, gas in the 0.3-1 $R_{\rm vir}$ bin in the left panel exhibits cooling within a 1 $h^{-1}$Mpc radius to a filament, in line with the radius of influence exhibited for colour (Fig.~\ref{fig:colcombo}) and cold gas fraction (Fig.~\ref{fig:cgfcombo}). 

Note that in both panels, gas deeper within a cluster (orange to blue) is consistently hotter across all distances to filament and angular separations. This is expected as the temperature of the diffuse gas is demonstrated to increase deep within clusters due to the fact that density also increases. Note that while radiative processes are observed to reverse this trend in the ultra-dense, innermost cores of "cool core" clusters, this phenomenon occurs on scales significantly smaller than the minimal averaging scale probed in this study ($0.3\, R_{\rm 100})$). However, cluster gas within $0.3-1 R_{\rm vir}$ at small filament angular separations ($D_{\rm fil}/D_{\rm cent}<0.2$) is actually colder than outskirt gas (within $1-2 R_{\rm vir}$) at large separations ($D_{\rm fil}/D_{\rm cent}>0.4$), a region where galaxies typically undergo little to no quenching.

Temperature plateaus/monotonously increases for the deepest gas bin. As discussed for the density profile, at this depth moving towards the filament highly correlates to moving towards the cluster centre. Thus, this rise in temperature is due to the fact that density increases towards the cluster centre. 

Since both halos (Section~\ref{sec:gasfracan}) and diffuse material (this Section) exhibit a stark increase in cold gas within the inner 1 $h^{-1}$.Mpc of filaments, and progressively with decreasing angular separation,  this suggests that interactions between halos and the intra-filament medium might be the origin of the resistance of local halos to quenching. In the following section, we examine how both the diffuse gas and halos flow along intra-cluster filaments.

\subsection{Flows near filaments}
\label{sec:galflow}
To understand how the cooler gas streams of intra-cluster filaments interact with halos, we first examine their respective flows along filaments. 

\begin{figure*}
	\includegraphics[width=\columnwidth]{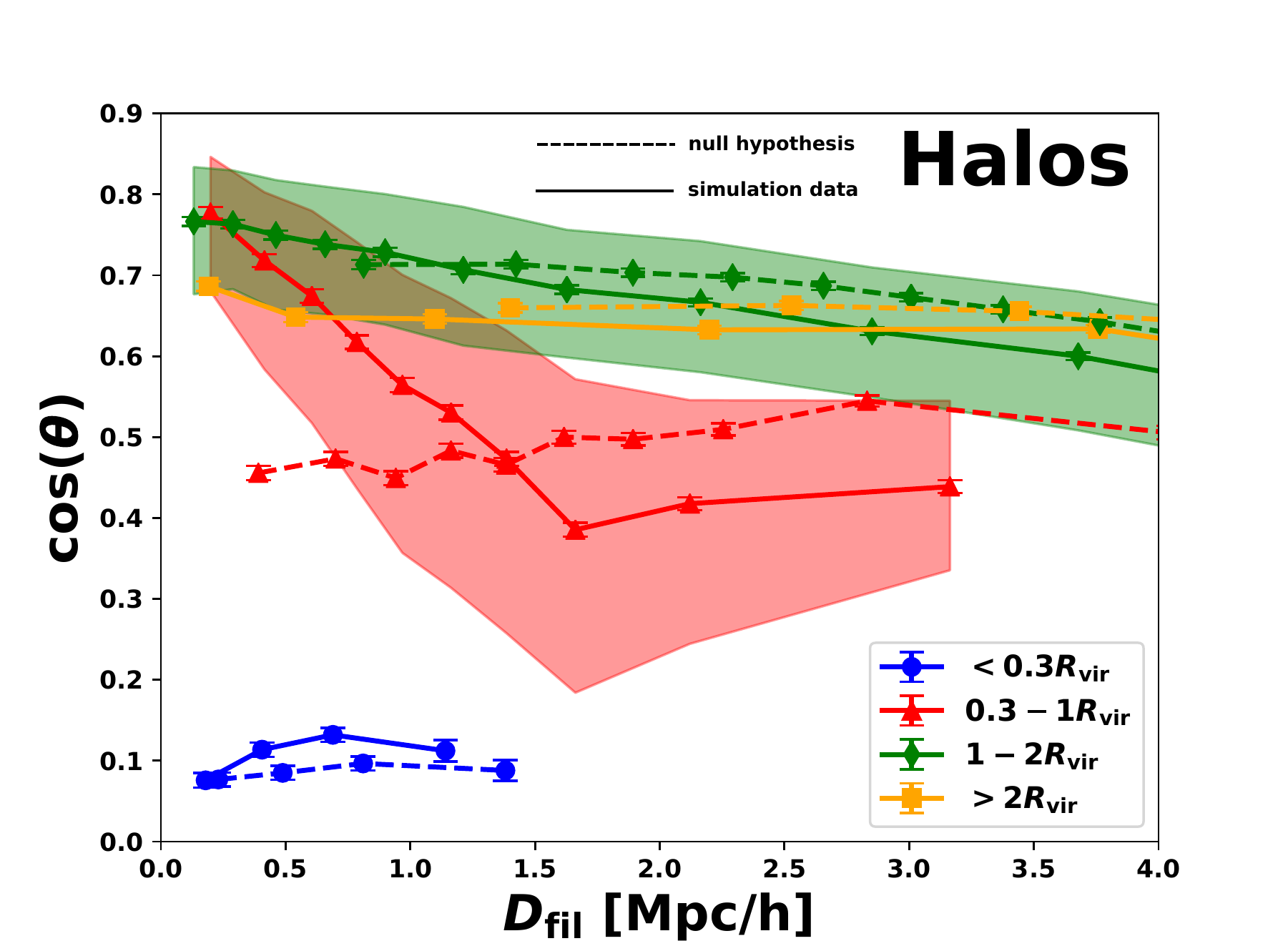}
	\includegraphics[width=\columnwidth]{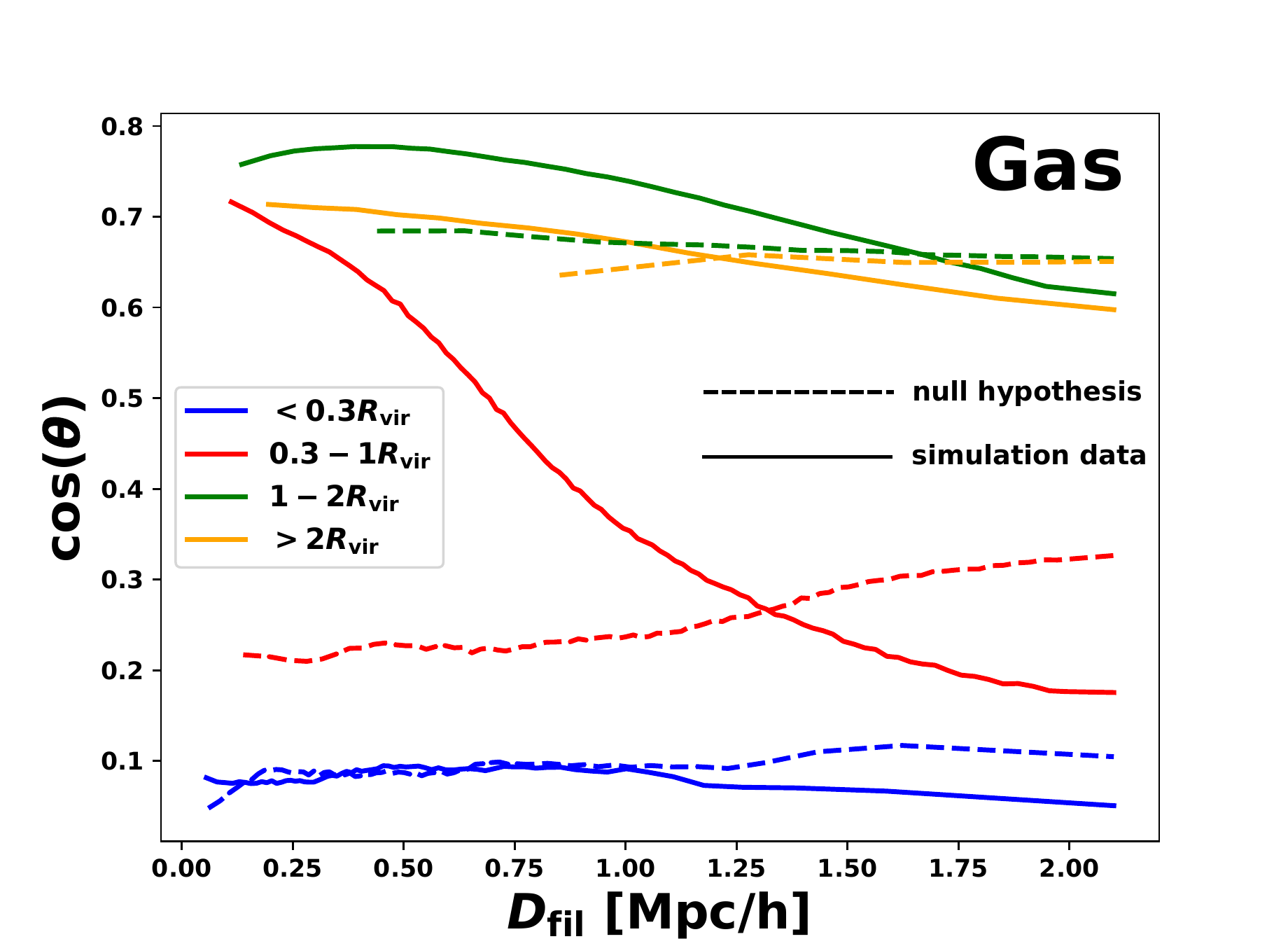}
    \caption{\textbf{Left Panel:} Median evolution of the cosine of the angle between a halo's velocity and its nearest filament versus $D_{\rm fil}$. Error bars are standard error on the mean and coloured contours indicate 40 to 60 percentiles.These latter are omitted below 0.3 $R_{\rm vir}$ for visibility. Dashed lines represent the null hypothesis. Halo flows between 0.3 and 1 $R_{\rm vir}$ align with filaments up to 2 $h^{-1}$.Mpc from the spine. \textbf{Right Panel:} Median evolution of the cosine of the angle between the diffuse gas velocity   and the nearest filament versus $D_{\rm fil}$. All errors on the mean are below 0.025 and are therefore omitted. Gas flows within the cluster strongly align with their filaments in their vicinity.}
    \label{fig:coscombo}
\end{figure*}

In the left panel of  Fig.~\ref{fig:coscombo}, we plot the median evolution of the cosine of $\theta$, the angle between the velocity vector of a halo and its nearest filament direction, as computed in Eq.~\ref{eq:cos} (oriented inwards). (The angular separation version of this plot can be found in Appendix~\ref{sec:cosdfdcapp}.) Halos are split into bins by their cluster-centric distance. The dashed lines represents the null hypothesis (no specific impact of filaments, hence roughly isotropic flows into the cluster ) whereby the filament network of each cluster is rotated by $45^{\circ}$ in a random direction before the measurements as explained in Section~\ref{sec:flows}. A median value of $\cos(\theta)=1$ means all halos are flowing inwards parallel to their nearest filament, and in  case of random flows we expect a median $\cos(\theta)=0$. In theory, values close to 0 can also be obtained if all halos flow perpendicular to the filament but in such a case we would also have a median $|\cos(\theta)|\approx 0$ and we verified that $|\cos(\theta)|>0.5$ in all bins up to 4 $h^{-1}$.Mpc away from filaments, suggesting at least relative alignment due to radial flows into the cluster. 

Error bars indicate standard error on the mean, while coloured contours indicate the spread from the 40th to the 60th percentile. There are 5000 halos per bin.

At distances to filament ($D_{\rm fil}$) greater than 1 $h^{-1}$.Mpc, halos residing beyond 1 cluster virial radius (1-2 and >2 $R_{\rm vir}$) have generally high $\cos(\theta)$ values, above 0.6, higher than those residing inside 1 virial radius (<0.3 and 0.3-1 $R_{\rm vir}$). This indicates that, within 4 $h^{-1}$.Mpc from the spine, halos outside the cluster environment tend to flow more parallel to their filaments and more consistently inwards than halos inside the cluster environment, which evolve in a highly mixed, multi-streaming region and can also experience backsplash. However, the specific boost in coherence at the spine of filaments at $D_{\rm cent}> 1 R_{\rm vir}$ (i.e. not merely attributable to steady radial flows into the cluster) is somewhat limited  ($\approx$ 0.05 difference in $\cos(\theta)$ with the null hypothesis (isotropic cluster flows) at the spine).

In contrast, halos 0.3 to 1 $R_{\rm vir}$ away from the cluster centre display a strong steady increase in median $\cos(\theta)$ within 2 $h^{-1}$.Mpc from the filaments, from 0.4 to 0.8. This indicates that galaxy flows go from less to markedly more aligned with intra-cluster filaments closer to the spine. It is important to notice that this sharp increase in $\cos(\theta)$  is not replicated in the null hypothesis for these haloes, indicating that it is not simply a geometric correlation with predominantly radial flows. In fact, galaxy flows start to align more with filaments than the isotropised expectation at 1-1.5 $h^{-1}$.Mpc from the spine,  in line with the previously estimated filament radius of influence on cold gas fraction.

Moving to the cluster's core, halos closest to the centre ($<0.3 R_{\rm vir}$) have $\cos(\theta)$ values close to 0 across the filamentary distance range they are found. This indicates high mixing and little to no correlation between the flow of these halos and their nearest filament. In line with our discussion in the previous section, we do not expect filaments to stand out as cold gas flows in this regime.

We next look at how the diffuse gas (i.e not in satellites) is flowing around and away from filaments using the same measure. The right panel of Fig.~\ref{fig:coscombo} displays the evolution of median $\cos(\theta)$, computed from the gas particles' velocities, against $D_{\rm fil}$. (The angular separation version of this plot can be found in Appendix~\ref{sec:cosdfdcapp}.) Similar to halos, gas beyond the cluster virial radius (orange and green curves) tends to have higher $\cos(\theta)$ values than gas inside of the cluster virial radius (blue and red curves). This indicates that flows of gas outside of the cluster environment are strongly radial to the cluster and coherent along filaments even 2 $h^{-1}$.Mpc away from filaments. The specific enhancement of alignment with filaments (compared to the null hypothesis in dashed lines) is nonetheless clear within 1.5 $h^{-1}$.Mpc from the nearest filament.

Gas in the 0.3-1 $R_{\rm vir}$ bin sees a strong, progressive increase in $\cos(\theta)$ values at distances lower than 1.3 $h^{-1}$.Mpc from intra-cluster filaments, going from 0.2 to 0.7. The null hypothesis in this bin clearly does not exhibit this same increase, indicating that gas streams are markedly more coherent along intra-cluster filaments than further away and that this is not simply a geometric correlation with radial cluster flows.

Once again, gas closest to the cluster centre (blue curve) exhibits the lowest $\cos(\theta)$ values with fluctuation of less than 0.05 across the filamentary distance range. As with the halos in this regime, we are seeing mostly random flows, strengthening the conclusion that cool gas filaments do not persist at this cluster depth.

In summary, inside the cluster and down to 0.3 $R_{\rm vir}$, both gas and halos hosting galaxies are flowing together in coherent streams along filaments towards the cluster centre. This supports the idea that cluster galaxies close to filaments may retain the ability to accrete gas efficiently, and/or undergo reduced ram pressure stripping. This is further investigated in the next sections.

\subsection{Fraction of accreting  halos}
\label{sec:accfracan}

Given that halos and cooled gas flow together along intra-cluster filaments, we now analyse the ability of cluster halos to keep accreting cold gas near and far from filaments. We quantify how galaxies are accreting cold gas using Eqs.~\ref{eq:outflow} and \ref{eq:accfrac}. 

\begin{figure}
	\includegraphics[width=\columnwidth]{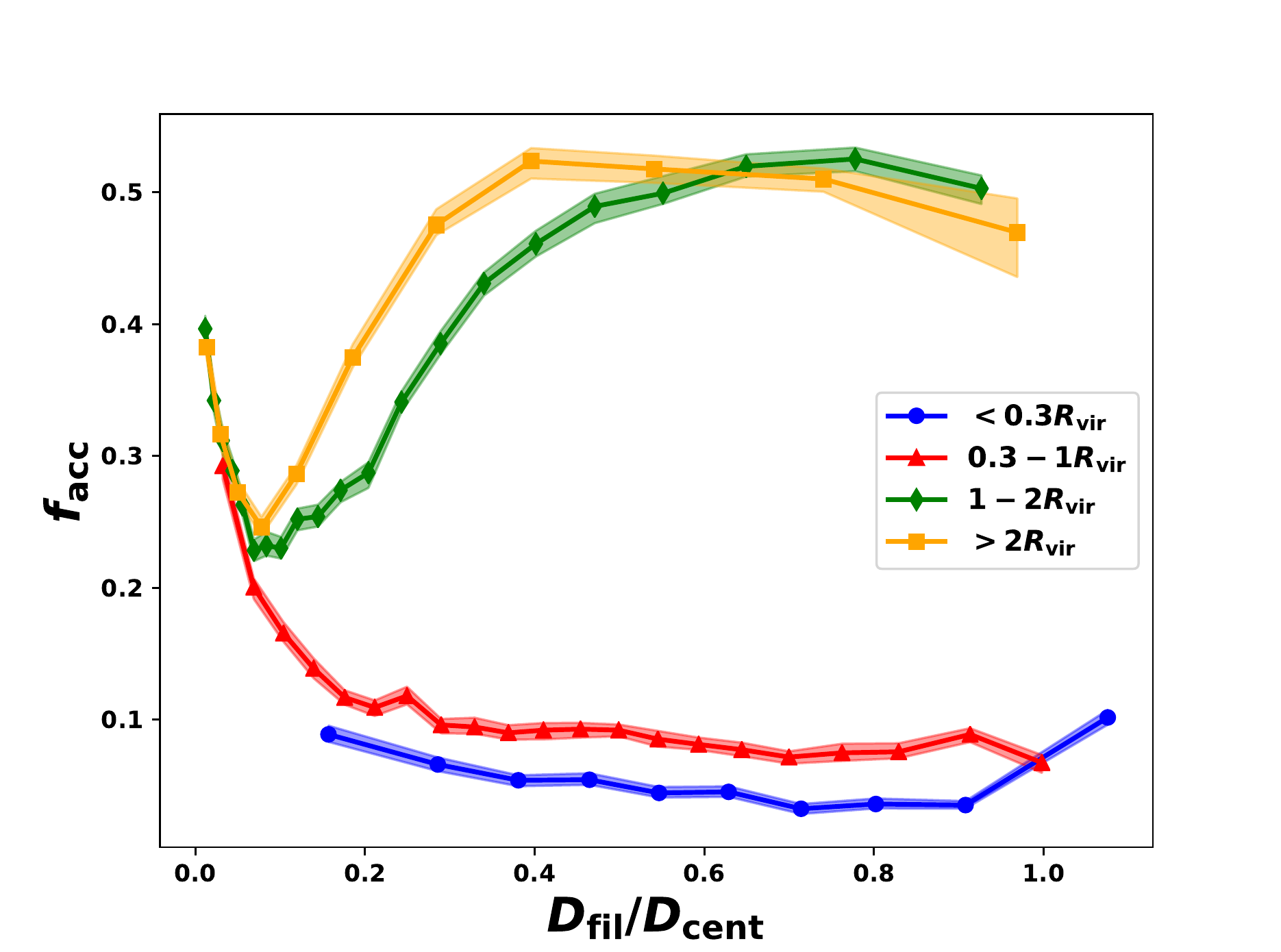}
    \caption{Fraction of halos accreting cold gas ($f_{\rm acc}$) versus angular separation to filament. Halos are binned by cluster centric distance. Coloured contours show 1-$\sigma$ bootstrap errors. $f_{\rm acc}$ increases at low angular separations for halos within 0.3-1 $R_{\rm vir}$  while it mostly decreases towards filaments outside clusters.}
    \label{fig:accvdfdc}
\end{figure}

Fig.~\ref{fig:accvdfdc} shows the fraction of halos accreting cold (T<10$^5$ K) gas against their angular separation ($D_{\rm fil}/D_{\rm cent}$) to their nearest filament. As usual, halos are binned based on their cluster-centric distance. Coloured contours indicate 1-$\sigma$ bootstrap errors. There are 2500 halos per bin.

As expected, halos residing beyond one virial radius (1-2 and >2 $R_{\rm vir}$) have consistently higher accreting fractions than those residing inside one virial radius (<0.3 and 0.3-1 $R_{\rm vir}$), across the entire angular separation range. Given that the field environment is colder and less turbulent than the cluster environment, it follows that it is generally more likely for a galaxy to be accreting cold gas in the field/outskirts of the cluster than inside the cluster itself.

These outskirt/field halos mostly experience a progressive decrease in accreting fraction at smaller angular separations. The secondary increase of accreting fraction at  the spine of filaments ($D_{\rm fil}/D_{\rm cent}<0.1$) is a bias, once again due to the presence of massive groups (hosting massive centrals) at the spine of filaments (Appendix~\ref{sec:mstevolapp}). For these halos outside the cluster, their accretion essentially decreases near filaments, as they undergo more interactions and experience pre-processing. 

In contrast, halos between 0.3 and 1 virial radius exhibit a progressive increase in accreting fraction closer to filaments, from below $10\%$ to $30\%$  from an angular separation of 0.3 to 0.1. In appendix~\ref{sec:mstctlapp} we show that this increase in accreting fraction at low angular separations is observed at all stellar masses and not simply related to the underlying mass distribution near filaments. Note that this also extends to evolution with halo mass.

At angular separations $D_{\rm fil}/D_{\rm cent}>0.3$, only around 10\% of halos in this cluster-centric distance range are accreting. This indicates that the coherent flows of gas and halos along filaments result in effective accretion of cold gas into filament halos inside the cluster environment.

No more than 10\% of halos within 0.3 $R_{\rm vir}$ are accreting at any given angular separation. As we have established before, this regime does not contain filaments in the form of cool gas flows, and the centre of the cluster environment contains the warmest gas as established in Fig.~\ref{fig:tempcombo}.

Within the cluster environment, these are clear indicators that the cooler gas flows that galaxies are streaming along with near filaments are refreshing their fuel supply in an otherwise extremely hot, violent environment. This supports the idea that intra-cluster filaments provide relief from quenching by reducing local strangulation.

\subsection{Ram pressure stripping}
\label{sec:rpsan}

Given that it is a dominant effect deep within clusters, we quantify the ram pressure stripping that our halos experience near and far from filaments to assess whether filaments can in fact reduce this effect.  We estimate ram pressure,  $P_{\rm ram}=\rho_{\rm shell} (\delta v)^2$, as described in Section~\ref{sec:rps}.

Outside clusters, simulations suggest that ram pressure values are typically lower than $10^{-13.5}$ Pa$\cdot$h$^2$, while deep inside clusters, they can increase up to around $10^{-11}$ Pa$\cdot$h$^2$  \citep[e.g.][]{Tonnesen_2009,Tecce_2010, Marshall_2017}. 

\begin{figure}
	\includegraphics[width=\columnwidth]{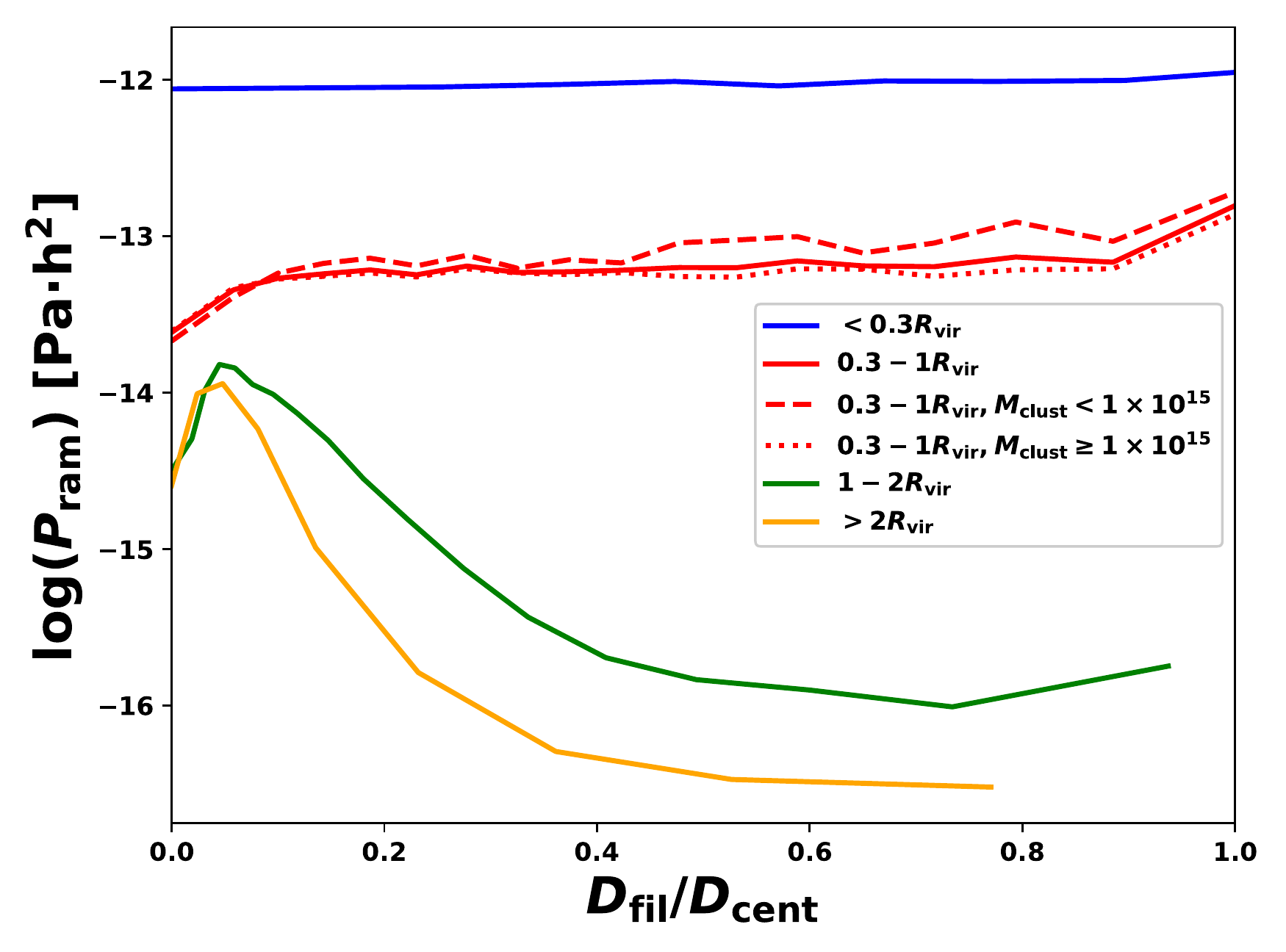}
    \caption{Evolution of median ram pressure with filament angular separation for halos which are binned by cluster-centric distance and host cluster mass (in units of $M_{\odot}$). In clusters, ram pressure slightly decreases at small angular separations .The opposite is true outside clusters.}
    \label{fig:rpsdfdc}
\end{figure}

Fig.~\ref{fig:rpsdfdc} displays the variation of ram pressure around halos with angular separation ($D_{\rm fil}/D_{\rm cent}$) to their nearest filament, separated by cluster-centric distance. Halos residing at 0.3-1 $R_{\rm vir}$ are further separated by the mass of the cluster in which they reside. 

Halos residing at 1-2 and >2 $R_{\rm vir}$ experience much lower ram pressure than those halos at <0.3 and 0.3-1 $R_{\rm vir}$. This is to be expected as halos typically enter the hot, dense, intra-cluster medium within a cluster's virial radius. We explore the evolution of ram pressure with cluster-centric distance in Appendix~\ref{sec:rpsdcentapp}.

For halos living at 1-2 and >2 $R_{\rm vir}$, ram pressure significantly increases from $10^{-16}$ or less to $10^{-14.5}$ Pa$\cdot$h$^2$ at decreasing angular separation to nearest filament ($D_{fil}/D_{cent}$ from 0.6 to 0.05) . The decrease noted at less than 0.1 angular separation from approximately $10^{-14}$ to $10^{-14.5}$ Pa$\cdot$h$^2$ is of little significance for stripping as it remains well below typical cluster values (as seen in Appendix~\ref{sec:rpsdcentapp}). The large increase in ram pressure approaching filaments in the field can largely be explained by the increase in density approaching field filaments in comparison to the background field.

Starkly different, halos living at <0.3 and 0.3-1 $R_{\rm vir}$ away from the cluster centre see no increase in ram pressure stripping with decreased angular separation to filaments, an effect mostly independent of cluster mass. Halos within 0.3-1 $R_{\rm vir}$ of clusters below $10^{15} M_{\odot}$ however experience a slight decrease in median ram pressure from around $10^{-12.9}$ Pa$\cdot$h$^2$ at angular separation 0.8 to around $10^{-13.2}$ Pa$\cdot$h$^2$ at angular separation 0.1 and even $10^{-13.6}$ Pa$\cdot$h$^2$ at the filament spine. This is the equivalent of median decrease in ram pressure a halo would experience being at 0.7 $R_{\rm vir}$ from the cluster centre rather than 0.5 $R_{\rm vir}$, as seen in  Fig.~\ref{fig:rpsdcentmed}, a significant difference in the context of quenching.

This is evidence that the coherent streams of cold gas along intra-cluster filaments are sufficient to compensate the filaments' limited contrast in density (compared to the ICM), hence halting the increase of ram pressure towards filaments and the stripping of gas from cluster galaxies.

Note that ram pressure stripping not only relies upon the ram pressure a galaxy experiences, but also on the gravitational potential of the galaxy. Given that halos are increasingly massive approaching the spine of filaments (Appendix~\ref{sec:mstevolapp}), it follows that these halos have larger gravitational potentials. This, in combination with the fact that ram pressure plateaus or even decreases approaching intra-cluster filaments, indicates that ram pressure stripping is actually significantly decreased for cluster galaxies near filaments. 

 We further analyse the dynamical extent of gas stripping on the halo scale in the next section.

\subsection{Quantifying the gas stripping in halos: the gas unbinding parameter}
\label{sec:bgan}

Where ram pressure stripping is dominant, galaxies have long tails of gas trailing behind them \citep{Ebeling_2014, Poggianti_2017}. Once effectively stripped, nearly all cold, star-forming gas is completely disassociated with the star and dark matter content of a halo. That is, the gas is unbound. The unbinding parameter, $B_g$, defined in Section~\ref{sec:bgdef} quantifies how stripped/disturbed the gas content in a given halo is. The lower the $B_g$ of a halo, the more associated its gas content is with the stellar and dark matter components, and thus the less stripped (and more bound) it is. For $B_g\gtrapprox2$, the gas is mostly unbound and increasing values relate to increasing differential velocity between the halo and the ICM.

\begin{figure}
	\includegraphics[width=\columnwidth]{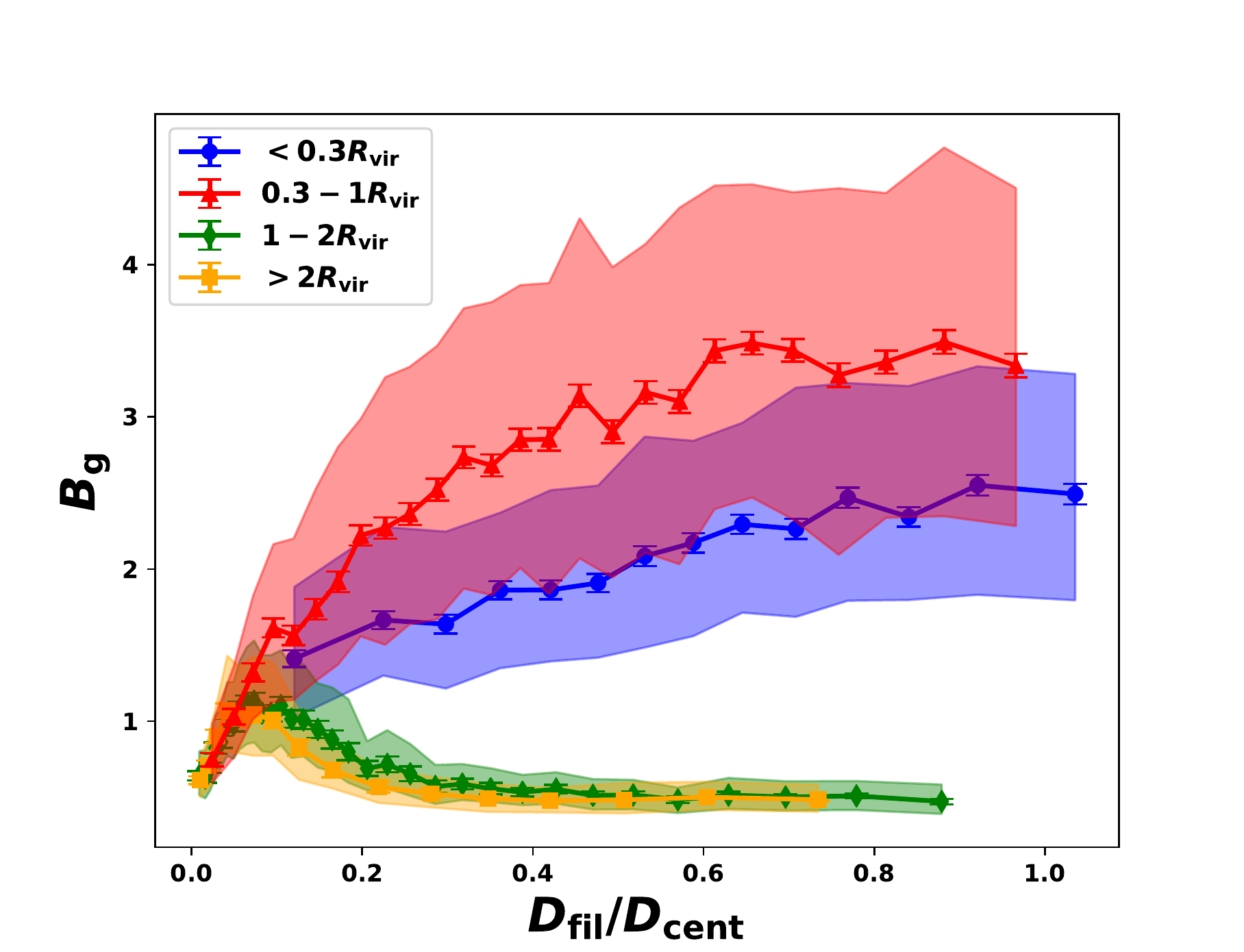}
    \caption{Median gas unbinding parameter ($B_g$) of halos versus filament angular separation, split by halo cluster-centric distance. Coloured contours indicate 40th to 60th percentiles, error bars are standard errors on the mean. Lower $B_g$ values indicate more bound gas. Cluster halos exhibit a smooth decrease in $B_g$ towards filaments, while outskirt halos mostly exhibit an increase.}
    \label{fig:cmvdfdc}
\end{figure}

Fig.~\ref{fig:cmvdfdc} shows the median evolution of $B_g$ with respect to a halo's angular separation ($D_{\rm fil}/D_{\rm cent}$) to its nearest filament. Halos are binned by their cluster-centric distance and coloured contours indicated the 40th to 60th percentile spread, while error bars indicate the standard error on the mean. There are 1500 halos per bin. 

Halos that reside at 1-2 and >2 $R_{\rm vir}$ have significantly lower $B_g$ values than those that reside at <0.3 and 0.3-1 $R_{\rm vir}$ when further than 0.2 angular separation from their nearest filament. As seen with other measures in this work, this is expected as galaxies in the field are in a quiet environment and thus retain their gas, while those in clusters are at various stages of being stripped. 

Halos residing at 1-2 and >2 $R_{\rm vir}$ see an increase in their $B_g$ value at smaller filament angular separations, particularly within $D_{\rm fil}/D_{\rm cent}<0.2$ . This indicates that galaxies closer to field filaments see an increased disassociation from their gas content, with tails of gas streaming behind them. This is consistent with the increased density and ram pressure stripping established in outskirt filaments. Note that the sudden decrease of $B_g$ at the spine ($D_{\rm fil}/D_{\rm cent}<0.05$) is once again a mass effect related to the local over-representation of massive groups less disturbed by the ICM (see Appendix~\ref{sec:mstevolapp}).  

In clear contrast with this result, halos residing 0.3 to 1 $R_{\rm vir}$ from the cluster centre display a strong, progressive decrease in their $B_g$ values at decreasing angular separation to their nearest filament, progressing from a median $B_g$ of over 3 at $D_{\rm fil}/D_{\rm cent}>0.5$ to below 1 at $D_{\rm fil}/D_{\rm cent}<0.1$. Cluster galaxies far from filaments are expected to be heavily stripped as they pass through the intra-cluster medium, resulting in high $B_g$ values reflecting gas mostly completely disassociated from the stellar and dark matter components of the halo. The decreased $B_g$ values of halos closer to intra-cluster filaments indicates that the gas in such halos is significantly less disturbed than that of halos further away from filaments. Hence such filaments are indeed providing galaxies relief from gas stripping, resulting in more bound gas in these galaxies.

 In Appendix~\ref{sec:mstctlapp}, we further verify that these results hold at all stellar masses. We also verify that similar trends are obtained for the total fraction of dry halos, which include both halos with $B_{\rm g}>2$ and halos with no gas at all. In Appendix~\ref{sec:mhalo}, we also check for any evolution with halo mass. These trends in gas binding are therefore distinctly correlated with the presence of gaseous filaments.

Note that virtually all satellites within 0.3 $R_{\rm vir}$ from their cluster centre have $B_g$ values greater than 1 but usually less than 3. Satellites in this region are significantly stripped of their gas, yet their differential velocity with the ICM is much reduced as they lie in the cluster centre.

Overall, we have demonstrated that the steady supply of increasingly coherent, colder gas flows not impeded by ram pressure stripping in filaments limits the environmental quenching of galaxies, instead sustained by effective accretion, higher gas fractions and increased star formation rates. This effect is detected at all stellar and halo masses.

\section{Discussion and Conclusion}
\label{sec:summary}
In this study, we used 324 simulated clusters to analyse how the properties of cluster galaxies vary with respect to their proximity to cosmic filaments. We find that galaxies exhibit delayed quenching closer to intra-cluster filaments, in contrast to the established trends for galaxies near field or outskirt filaments. We find decreased strangulation and ram pressure in the vicinity of intra-cluster filaments to be the main causes of these trends.

In detail, the trends we find for galaxies are that:
\begin{itemize}
\item The fraction of star forming cluster galaxies increases closer to filaments, irrespective of stellar mass. This indicates that intra-cluster filaments actively modulate star formation and are not just hosting more massive galaxies than their surroundings.
\item Outside clusters, galaxies are redder closer to filaments. Inside clusters, they level off in colour and are bluer closer to filaments. 
\item Outside clusters cold gas fraction decreases for galaxies closer to filaments. Inside clusters, satellites near filaments display significantly increased cold gas fractions compared to their counterpart satellites further from filaments.  
\end{itemize}

These results show that, while filaments are associated with pre-processing in low or average density environments, deep cluster filaments delay the quenching of galaxies in the hot, dense cluster environment. Further investigating the origin of this quenching delay we find that:
\begin{itemize}
\item Filaments diving deep into clusters, down to $0.3 R_{\rm vir}$, are regions of coherent gas flows markedly colder than the ICM.
\item The density contrast of intra-cluster filaments with the intra-cluster medium is lower than for field or outskirt filaments with their background environments.
\item Galaxies and diffuse gas flow together along filaments. This is enhanced closer to their spines.
\item The fraction of galaxies accreting cold gas within the general cluster environment dramatically increases approaching filaments, indicating a decrease in strangulation.
\item Ram pressure slightly decreases closer to intra-cluster filaments.  In addition, galaxy mass and thus gravitational binding tends to increase. As a consequence, ram pressure stripping is decreased for cluster galaxies closer to filaments.
\item The gas in galaxies closer to intra-cluster filaments is more bound and less disturbed than for galaxies in the rest of the ICM.
\end{itemize}

Intra-cluster filaments are therefore clearly a region of reduced quenching where cluster galaxies can keep forming stars as they fall deeper into the cluster, a striking difference from their field and outskirt counterparts.

Further work will focus on investigating the time evolution of halos in the simulation, tracking halos as they evolve with the filaments to identify how quickly and strongly the delay in quenching appears. Higher resolution cluster simulations such as RomulusC \citep{tremmel} which better resolve the cluster regime, where hydrodynamic streaming instabilities and feedback processes may affect the structure of intra-cluster filaments, could also be used to confirm our analysis. DisPerSE has been used on data from large surveys such as GAMA \citep{Kraljic_2018, welker} and COSMOS \citep{Laigle_2017} to build the cosmic web. This can in principle be used to detect filaments plunging into clusters in 2D. This could be combined with detailed information on galaxy colour and gas content to confirm and extend our results in observations.

\section*{Acknowledgements}
SK and JW acknowledge the support of the Natural Sciences and Engineering Research Council of Canada (NSERC). CW acknowledges the support of the William and Caroline Herschel Fellowship program at McMaster University and the support of the National Science Foundation award 1815251 (United States) held by Dr. Susan Kassin. ZZ acknowledges the funding support of the Office of International Cooperation in USTC (University of Science and Technology of China). The analysis used publicly available software packages including Matplotlib (Hunter 2007) and NumPy (van der Wel 2008). This work was made possible by the `The Three Hundred' collaboration, with financial support from the European Union's Horizon 2020 Research and Innovation programme under the Marie Sklodowskaw-Curie grant agreement 734374 (LACEGAL project). GY acknowleges financial suport from  MICIU/FEDER through research grant number PGC2018-094975-C21. The simulations used in this paper have been performed in the MareNostrum Supercomputer at the Barcelona Supercomputing Center, thanks to CPU time granted by the Red Espa\~nola de Supercomputaci\'on. {\it Main authors contributed to this paper in the following ways: SK and CW share first-authorship. They contributed equally to the analysis, generated most plots and wrote the paper. ZZ generated colours and colour plots.  CW supervised SK and ZZ. JW provided scientific advice throughout the project. Dr. Pascal Elahi produced the halo catalogs. GY supplied the simulations. CW produced the filaments.} \\


\section{Data availability.}

The data underlying this article is available on request upon approval by The Three Hundred steering committee. Contact the corresponding author to request access.


\bibliographystyle{mnras}
\bibliography{kotecha20} 




\appendix

\section{Evolution of stellar mass with filament proximity}
\label{sec:mstevolapp}
\begin{figure}
	\centering
	\includegraphics[width=\columnwidth]{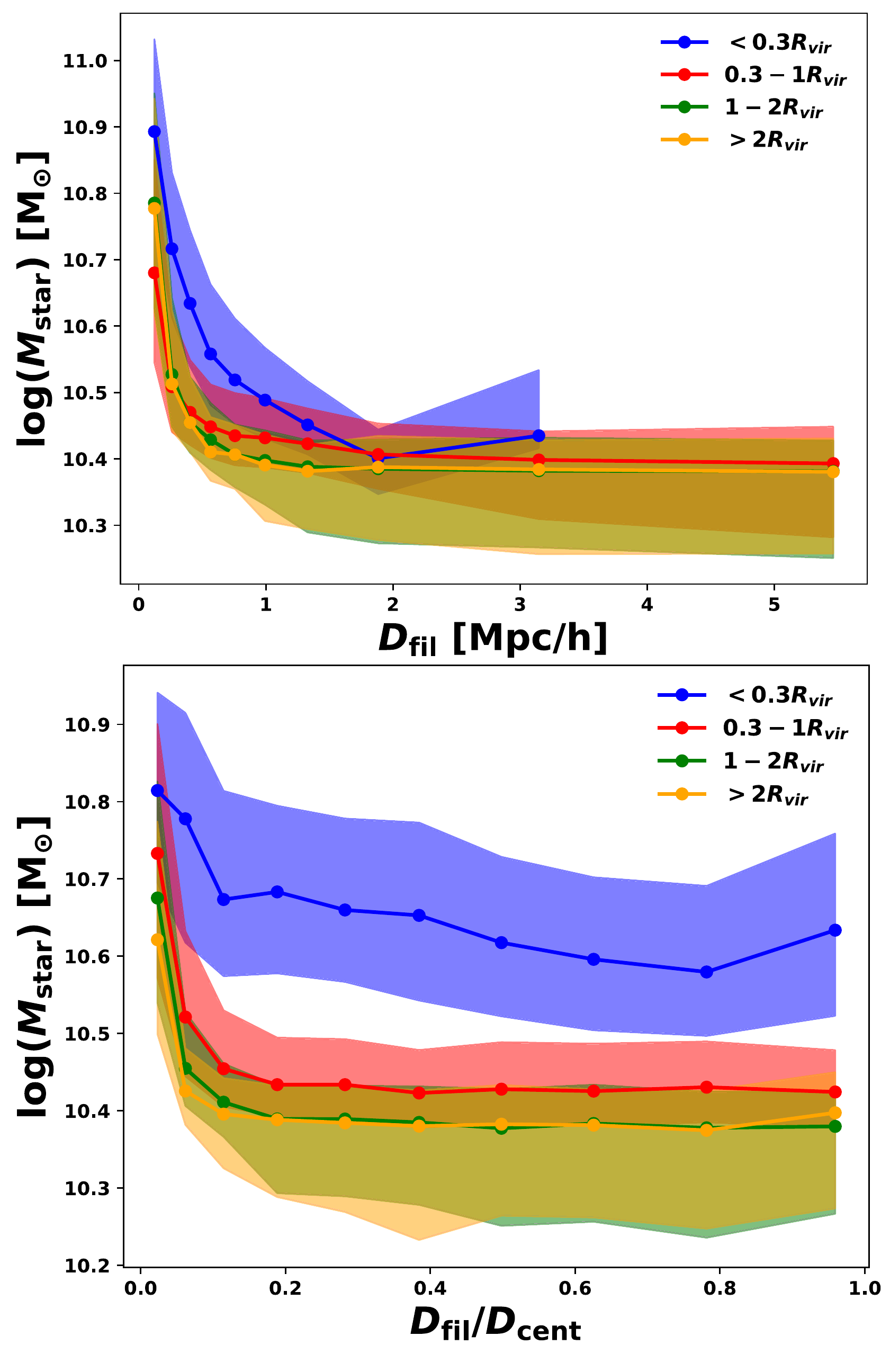}
    \caption{{\bf Upper panel: }Median halo stellar mass versus distance to nearest filament. {\bf Lower panel: }Median halo stellar mass versus angular separation to nearest filament. Coloured contours indicated 40-60 percentile spread. Stellar mass of halos increases approaching the spine of filaments.}
    \label{fig:mstdf}
\end{figure}

Fig.~\ref{fig:mstdf} shows the median halo stellar mass versus distance to nearest filament ($D_{\rm fil}$) and angular separation to nearest filament ($D_{\rm fil}/D_{\rm cent}$) respectively. Halos are separated by cluster-centric distance, and coloured contours indicate the 40th-60th percentile spread. 

Overall, as expected, stellar mass is increasing at lower distance to filament and angular separation but the increase is actually very peaked at the spine of filaments. Indeed, no change in median stellar mass is observed across cluster-centric bins further away than 1 $h^{-1}$.Mpc  from filaments (or at angular separation larger than 0.2). An increase in halo stellar mass is detectable at $D_{\rm fil}< 1$ $h^{-1}$.Mpc, but the most significant increase in stellar mass is confined within very low filament distances/angular separations ($\approx 80\%$ increase towards the spine across the inner 0.5 $h^{-1}$.Mpc or $D_{\rm fil}/D_{\rm cent}<0.05$).

It is notably clear that stellar mass is also increasing towards cluster centres, as inner halos (orange to green to red to blue) have higher stellar masses at fixed distance or angular separation to filament.
We also checked that median halo mass (DM mass) follows qualitatively similar trends with $D_{\rm fil}$ and $D_{\rm fil}/D_{\rm cent}$.

\section{Cosmic flow alignments versus angular separation}
\label{sec:cosdfdcapp}

\begin{figure}
    \includegraphics[width=\columnwidth]{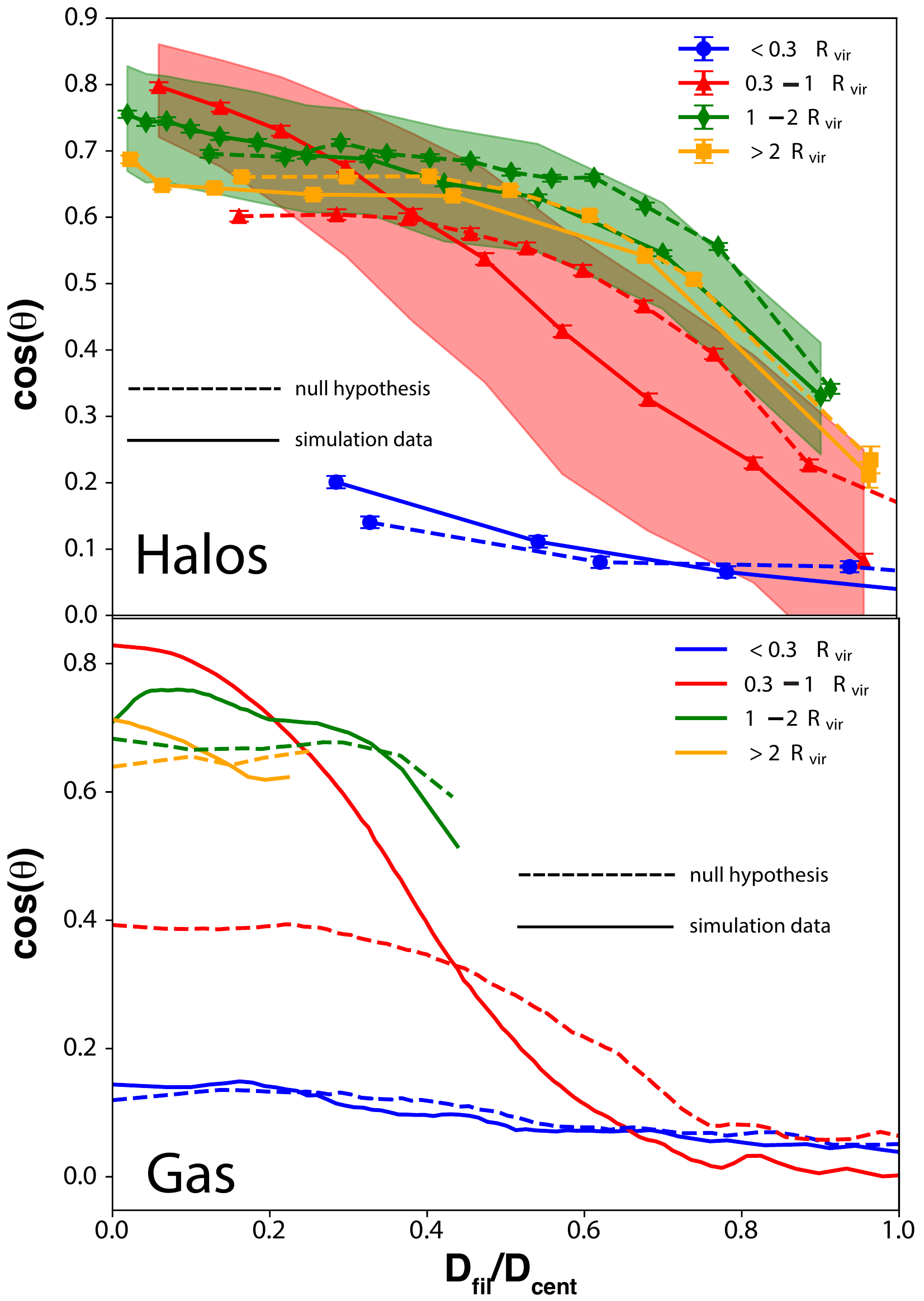}
    \caption{\textbf{Upper Panel:} Evolution of the median cosine of the angle between the velocity of a halo and its nearest filament versus angular separation to filament. Error bars are standard error on the mean, coloured contours are 40-60 percentiles. Dashed lines represent the null hypothesis. Halo flows between 0.3 and 1 $R_{\rm vir}$ become sharply aligned at the spine of filaments. \textbf{Lower Panel:} Same for gas particle velocities. As with halos, gas flows within the cluster become more aligned with their nearest filament.}
    \label{fig:cosdfdc}
\end{figure}

Fig.~\ref{fig:cosdfdc} shows the median evolution of the cosine of the angle between the velocity vector of halos (\textbf{upper panel}) and gas particles (\textbf{lower panel}) and their nearest filament versus angular separation ($D_{\rm fil}/D_{\rm cent}$). For the upper panel, error bars indicate standard error on the mean, coloured contours indicate the 40th-60th percentiles. Dashed lines stand for the null hypothesis (isotropic cluster flows). For both halos and gas particles between 0.3 and 1 $R_{\rm vir}$, there is a clear increase in cosine at smaller angular separations, beyond the mere geometric correlation between filaments and radial flows. This indicates that gas and halos flow effectively together along intra-cluster filaments.

Notably, separate analysis of $|\cos(\theta)|$ found that for halos outside their cluster virial radius (1-2 and >2 $R_{\rm vir}$), further than angular separations of 0.7, $|\cos(\theta)|<0.5$. For all other angular separations for halos at any cluster-centric distance, $|\cos(\theta)|\geq0.5$. This is in line with a picture where for field galaxies and filaments, matter is attracted perpendicularly at first towards filaments and then flows bend towards alignment with the filament in their vicinity.

\section{Evolution of fraction of accreting halos and gas unbinding parameter with stellar mass inside and outside the cluster}
\label{sec:mstctlapp}

\subsection{Fraction of accreting halos}

\begin{figure}
	\includegraphics[width=\columnwidth]{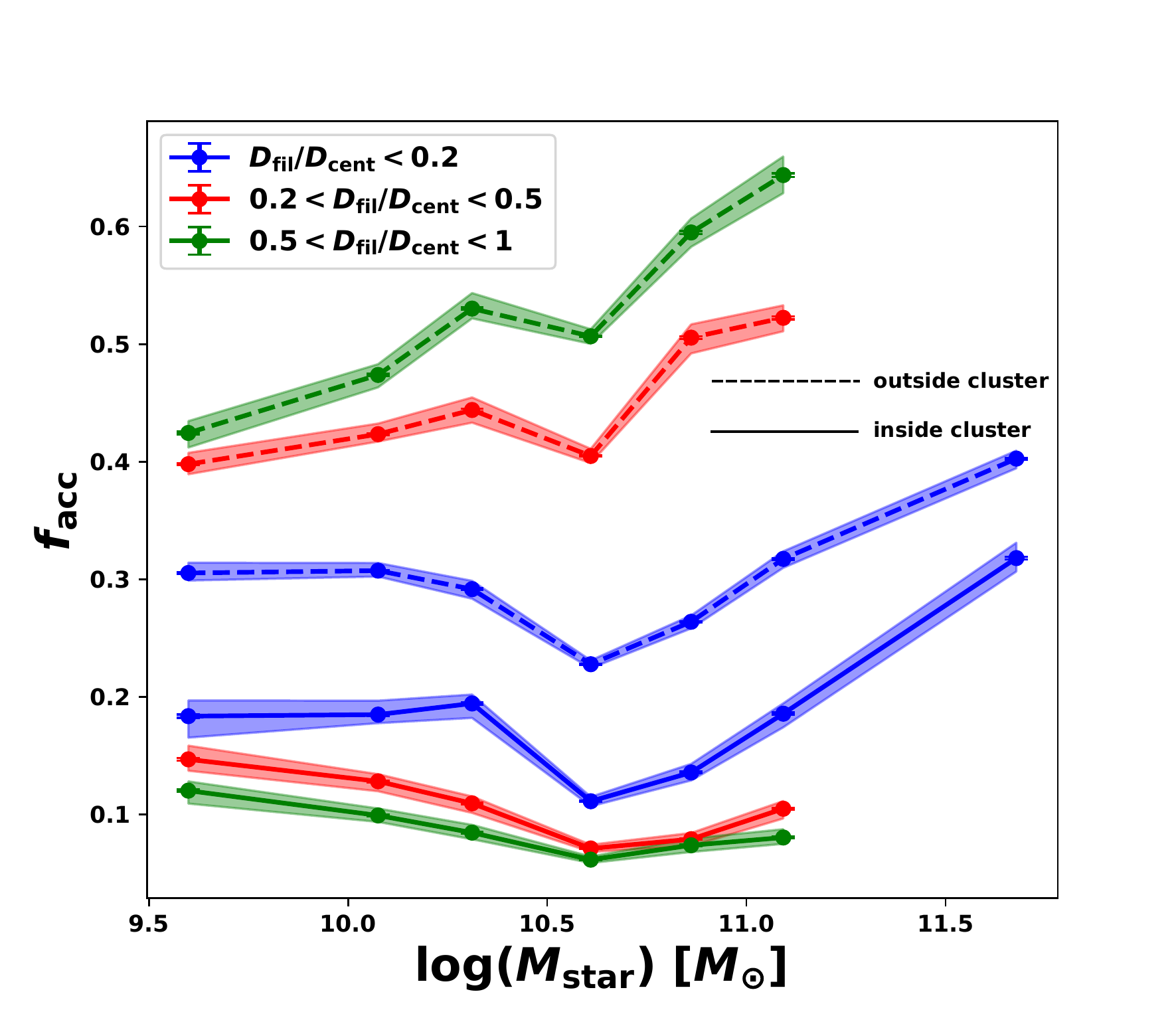}
    \caption{Fraction of accreting halos versus stellar mass, binned by angular separation to filament and residency within a cluster. Coloured contours show 1-$\sigma$ bootstrap errors. Outside clusters, $f_{\rm acc}$ decreases at smaller angular separations. Inside clusters, it conversely increases at smaller angular separations.}
    \label{fig:accmstctl}
\end{figure}

Fig.~\ref{fig:accmstctl} shows the fraction of halos accreting cold gas versus stellar mass, binned by their angular separation (colours) and residency inside (solid lines) or outside (dashed lines) the virial radius of clusters. Coloured contours indicate $1-\sigma$ bootstrap errors. As expected, the fraction of accreting halos is higher outside clusters than inside clusters at all angular separations. 

Outside clusters, at fixed stellar mass, the accreting fraction decreases at smaller angular separation (green to red to blue), consistent with pre-processing in field/outskirt filaments. 

Inside clusters, this trend reverses, and the accreting fraction increases at smaller angular separation, irrespective of stellar mass. This indicates that intra-cluster filaments are regions where halos are more readily able to accrete than in the rest of the cluster environment. 

Note also that most massive halos (hosting massive centrals $>10^{11.5} M_{\odot}$) lie in the smallest angular separation range (blue), and these tend to have higher accreting fractions. This $f_{\rm acc}$ trend with stellar mass in the high mass range is responsible for the apparent secondary increase of accreting fraction at the spine of filaments ($D_{\rm fil}/D_{\rm cent}<0.07$) seen in Fig.~\ref{fig:accvdfdc}.

\subsection{Gas unbinding parameter}

\begin{figure}
	\includegraphics[width=0.9\columnwidth]{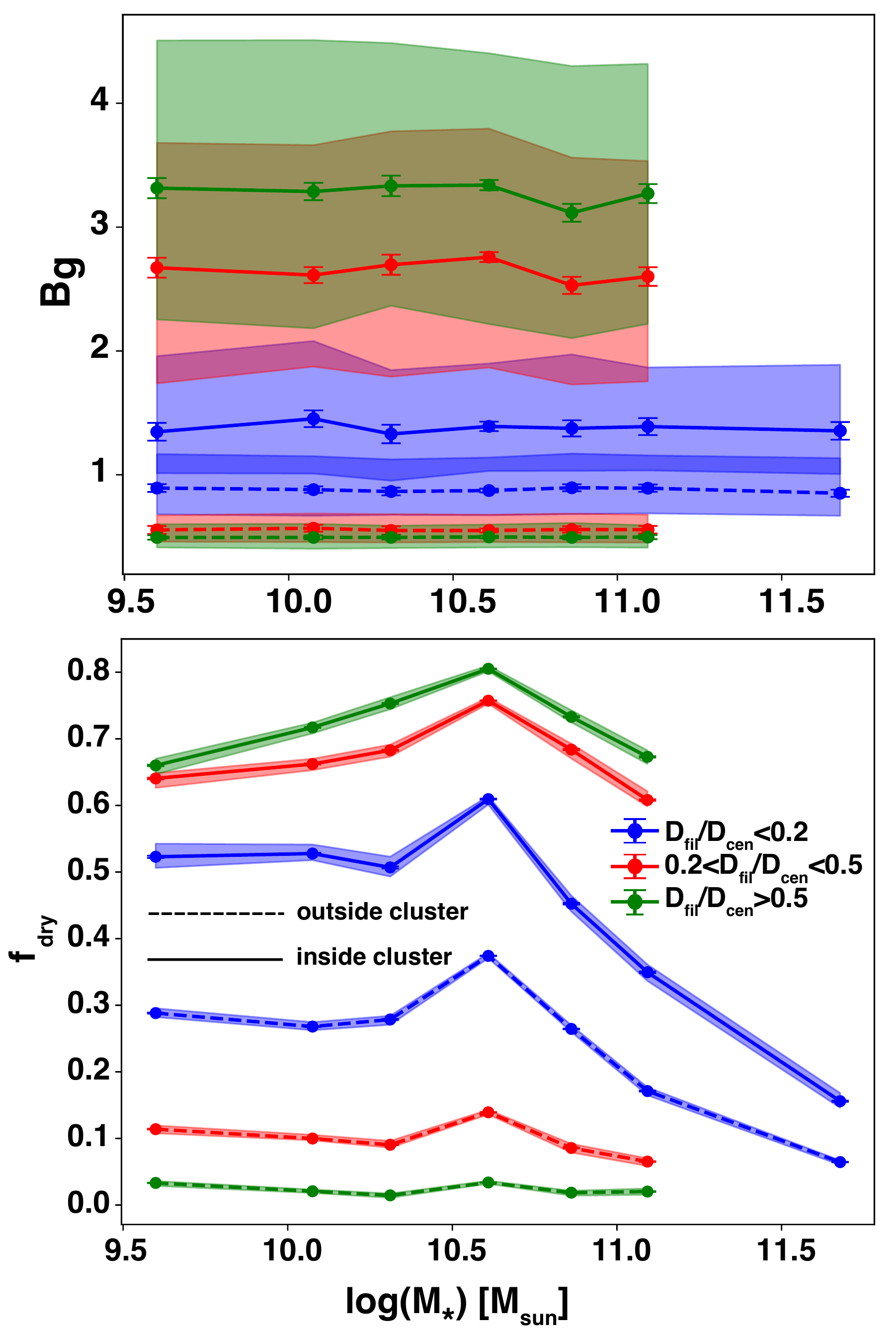}
    \caption{\textbf{Upper Panel:} Median $B_g$ versus halo stellar mass, binned by $D_{\rm fil}/D_{\rm cent}$ and residency within a cluster. Coloured contours are 40-60 percentiles. Outside clusters, $B_g$ increases at smaller angular separation while, inside clusters, it increases. \textbf{Lower Panel:} Fraction of dry halos versus halo stellar mass, with same binning. Coloured contours represent 1-$\sigma$ bootstrap errors. Outside clusters, dry fraction increases at smaller angular separation while it decreases inside clusters.}
    \label{fig:bgmstctl}
\end{figure}

Fig.~\ref{fig:bgmstctl}, upper panel, shows the median $B_g$ versus stellar mass while the lower panel displays the median fraction of dry halos $f_{\rm dry}$ (Eq.~\ref{eq:dryfrac}) versus stellar mass. Results are binned by angular separation and residency inside (solid) or outside (dashed) $R_{\rm vir}$ of clusters. Coloured contours indicate the 40th-60th percentiles (upper panel) or 1-$\sigma$ bootstrap errors (lower panel). Expectedly, halo gas is markedly less bound/more disturbed inside clusters than outside clusters at all angular separations, and halos are drier inside clusters than outside clusters, as demonstrated by larger $B_g$ and $f_{\rm dry}$ values. 

Outside clusters, $B_g$ remains mostly below 1 across the full range of stellar masses and angular separations. It is clear that, at fixed stellar mass, $B_g$ and $f_{\rm dry}$ increase at smaller angular separation (green to red to blue), consistent with pre-processing in field filaments causing more gas disturbance and loss. 

Inside clusters, this trend reverses: $B_g$ and $f_{\rm dry}$ decrease at smaller angular separations. This indicates halo gas is less disturbed and binds better to halos in intra-cluster filaments than in the rest of the ICM. Note that the most massive halos ($M>10^{11}M_{\odot}$) reside at the filament spine, and generally display very low dry fractions, again due to the fact that these are massive groups that seldom feel the impact of the ICM.

\section{Evolution of gas binding parameter with halo mass for genuine satellites.}
\label{sec:mhalo}

\begin{figure}
	\includegraphics[width=1\columnwidth]{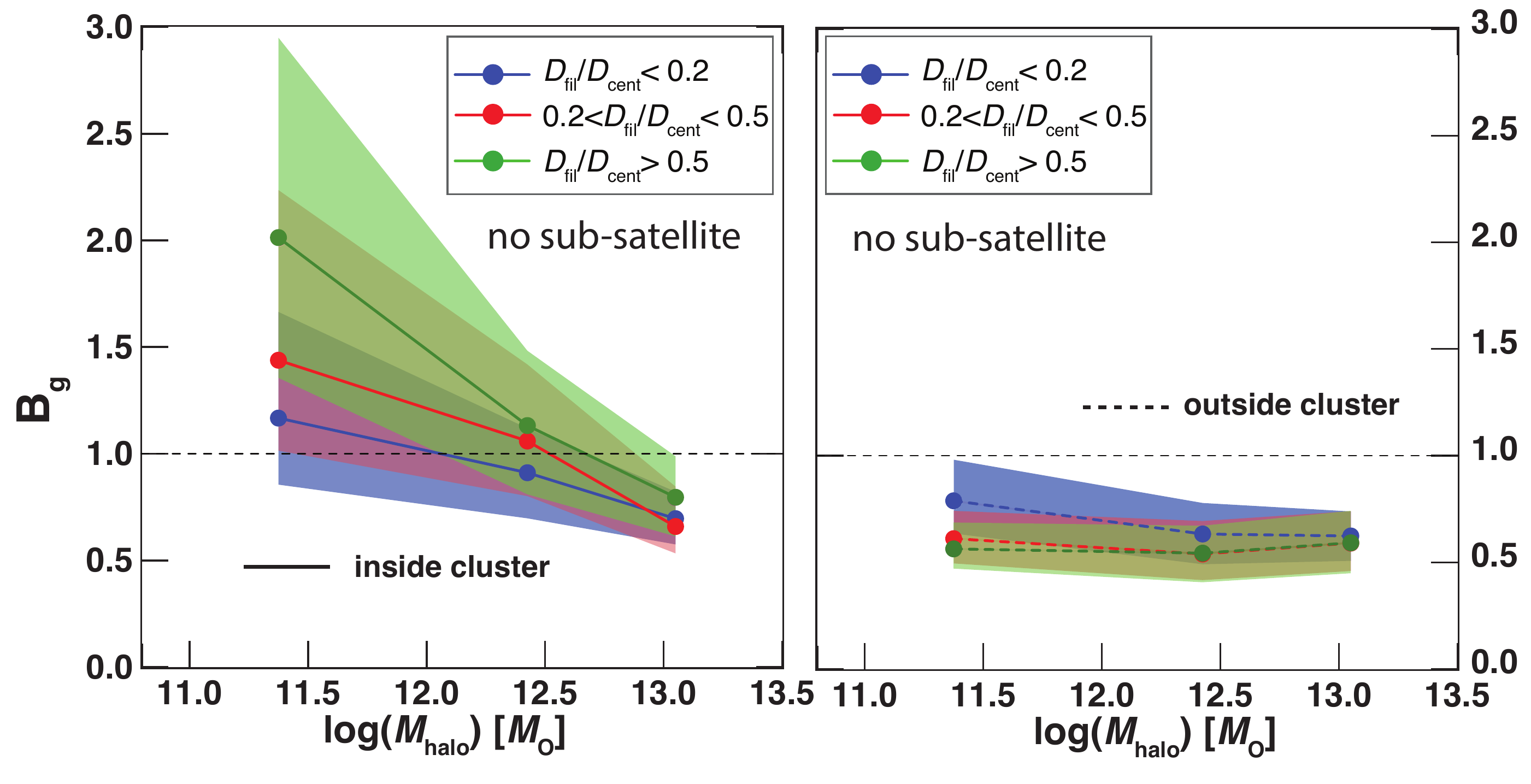}
    \caption{Median $B_{\rm g}$ versus halo mass for all genuine satellites inside clusters 
(left panel) and outside (right panel), for various angular separations from filaments (colours). Shaded areas show the 40th to 60th percentile. $B_{\rm g}$ (gas disturbance) increases inside clusters but halos closer to the filament experience a more limited increase.}
    \label{fig:mhalo-bg}
\end{figure}

Fig.~\ref{fig:mhalo-bg}, left panel, shows the median $B_g$ versus halo mass inside clusters, for all genuine cluster satellites (not satellite of a bigger satellite). The right panel displays the corresponding trend outside clusters ($>R_{\rm vir}$). Results are binned by angular separation (colours). Coloured contours indicate the 40th-60th percentiles (upper panel). Expectedly, halo gas is markedly less bound/more disturbed inside clusters than outside clusters at all angular separations as demonstrated by larger $B_g$.

Outside clusters, $B_g$ remains mostly below 1.0 across the full range of halo masses and angular separations. At fixed halo mass, it increases at smaller angular separation (green to red to blue), consistent with pre-processing in field filaments causing more gas disturbance and loss. 

Inside clusters, the reverse trend is observed, with halos at smaller angular separation experiencing reduced gas disturbance ($Bg<1.2$ across all halo masses for $D_{\rm fil}/D_{\rm cent}<0.2$). This confirms that filaments are able to reduce stripping independently of DM halo mass. It should be noted that most massive halos ($M_{\rm vir}>10^{13}\,M_{\odot}$) remain the most resistant to gas disturbance, with at least $60\%$ of them retaining $B_{\rm g}<1$ even at $D_{\rm fil}/D_{\rm cent}>0.5$.

\section{Evolution of ram pressure with distance to the centre of the cluster.}
\label{sec:rpsdcentapp}
\begin{figure}
	\includegraphics[width=\columnwidth]{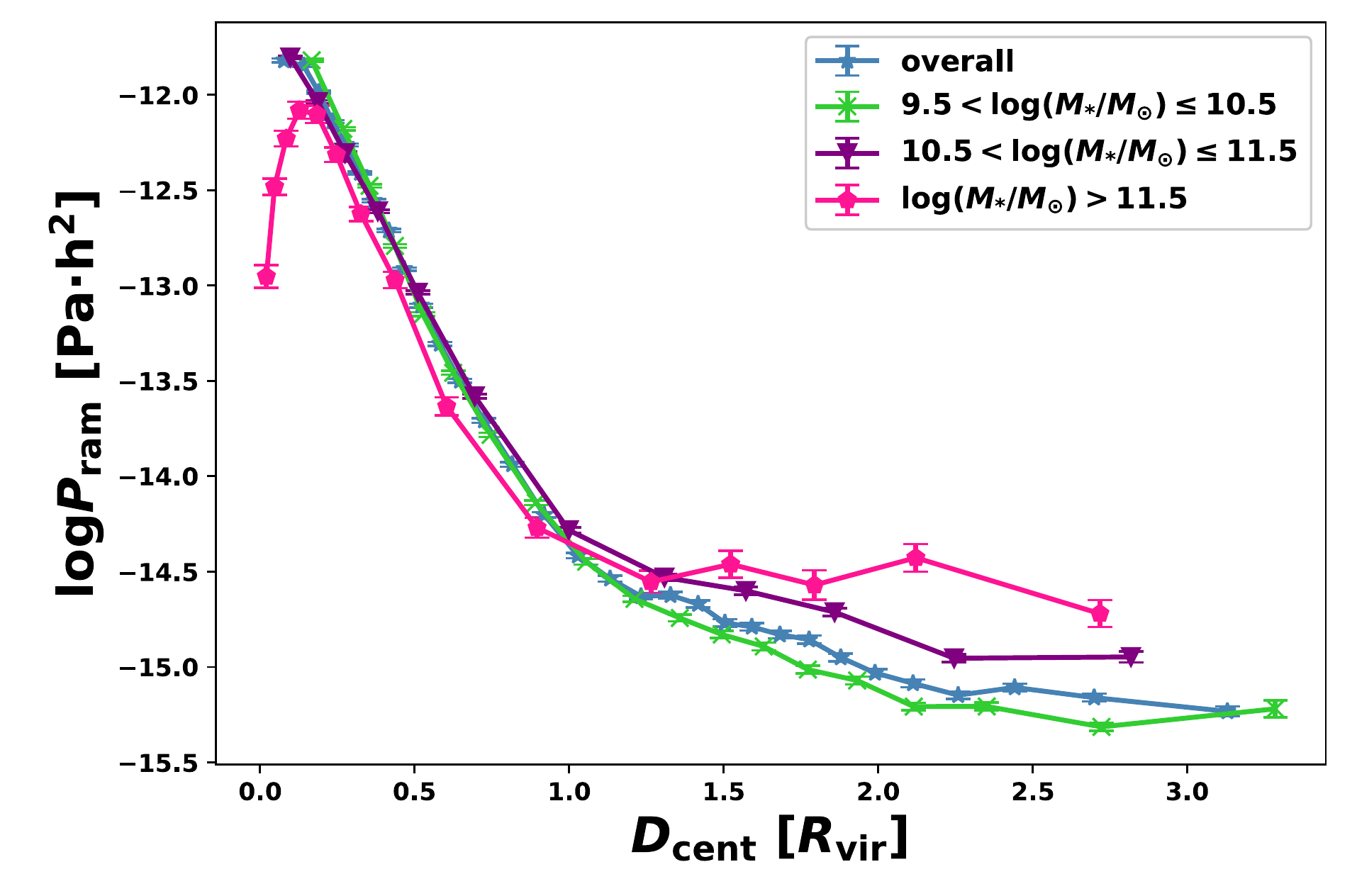}
    \caption{Median ram pressure exerted on halos against cluster-centric distance. Halos are binned by stellar mass. Error bars are standard errors on the mean. Ram pressure steadily increases for halos deeper within clusters.}
    \label{fig:rpsdcentmed}
\end{figure}

Fig.~\ref{fig:rpsdcentmed} shows the median ram pressure experienced by halos against their cluster-centric distance. Results are presented for the full sample (in blue) and for different stellar mass bins (green to pink). Error bars are standard error on the mean. 

For all stellar masses, ram pressure increases dramatically from approximately $10^{-14}$ Pa$\cdot$h$^2$ beyond 1 $R_{\rm vir}$ up to approximately $10^{-12}$ Pa$\cdot$h$^2$ at the cluster centre. Only most massive halos lying in the innermost core of the cluster ($D_{\rm cent} <0.2 R_{\rm vir}$) deviate from this trend as they have virtually no differential velocity with the host cluster. Most are cluster central candidates or the second brightest core galaxy.







\label{lastpage}
\end{document}